\documentclass{article}
\usepackage{eurosym}
\usepackage{amsmath}
\usepackage{graphicx}
\usepackage{float}
\usepackage[margin=1in]{geometry}

\graphicspath{ {./} }

\input{tcilatex}
\begin{document}

\begin{center}
\bigskip

\bigskip

\textbf{Beyond the classical Hamilton's rule: State distribution asymmetry
and the dynamics of altruism}\bigskip

\bigskip

Krzysztof Argasinski*,

Departement of Mathematics Informatics and Mechanics

University of Warsaw

ul. Stefana Banacha 2

02-097 Warszawa

Poland

\bigskip

Ryszard Rudnicki

Institute of Mathematics of Polish Academy of Sciences

ul. \'{S}niadeckich 8

00-656 Warszawa

Poland
\end{center}

*corresponding author: \textit{argas1@wp.pl}\bigskip

\textbf{Acknowledgments}: We would like to thank Mark Broom and Jan Koz\l 
owski for their support of the project and helpful suggestions and John
McNamara for hospitality during research stays and valuable advice and
Matthijs Van Veelen for helpful suggestions. This paper was supported by the
Polish National Science Centre Grant No.OPUS 2020/39/B/NZ8/03485 (KA) and received 2024 Tom Vincent Award by International Society for Dynamic Games.

\bigskip \bigskip \bigskip \bigskip

.\newpage

\bigskip

\textbf{Abstract}

This paper analyzes the relationships between demographic and state-based
evolutionary game framework and Hamilton's rule. It is
shown that the classical Hamilton's rule (counterfactual method), combined
with demographic payoff functions, leads to easily testable models. It works
well in the case when the roles of donor and receiver are randomly drawn
during each interaction event. This is illustrated by the alarm call
example. However, we can imagine situations in which role-switching results
from some external mechanism, for example, fluxes of individuals between the
border and the interior of the habitat, when only border individuals may
spot the threat and warn their neighbors. To cover these cases, a new model
is extended to the case with explicit dynamics of the role distributions
among carriers of different strategies, driven by some general\ mechanisms.
It is thereby shown that even in the case when fluxes between roles are
driven by selectively neutral mechanisms (acting in the same way on all
strategies), differences in mortality in the focal interaction lead to
different distributions of roles for different strategies. This leads to a
more complex rule for cooperation than the classical Hamilton's rule. In
addition to the classical cost and benefit components, the new rule contains
a third component weighted by the difference in proportions of the donors
among carriers of both strategies. Depending on the sign, this component can
be termed the \textquotedblleft survival surplus\textquotedblright , when
the donor's survival is greater than the receiver's survival, or the
\textquotedblleft sacrifice cost\textquotedblright\ (when it decreases the
benefit), when the receiver's survival exceeds that of the helping donor.
When we allow different role-switching rates for different strategies,
cooperators can win even in the case when the assortment mechanism is
inefficient (i.e., the probability of receiving help for noncooperators is
slightly greater than for cooperators), which is impossible in classical
Hamilton's rule.\bigskip

\newpage

\textbf{List of important symbols:}

\begin{tabular}{ll}
$n_{s}^{i}$ & number of individuals in state $i$ with strategy $s$ \\ 
$R_{s}^{i}$ & growth rate of individuals in state $i$ \\ 
& \quad with strategy $s$ \\ 
$\Lambda ^{i}$ & intensity of leaving state $i$ \\ 
$q_{s}^{i}=n_{s}^{i}/\sum_{j}n_{s}^{j}$ & frequency of individuals in state $
i$ among \\ 
& \quad individuals with strategy $s$ \\ 
$g_{s}$ & frequency of all $s$-strategists \\ 
$R_{b}$ & background growth rate \\ 
$d_{s}^{i}$ & mortality of individuals in state $i$ with strategy $s$ \\ 
& \quad lower index indicates strategy \\ 
& \quad $C$ -- cooperative $N$ -- noncooperative \\ 
& \quad upper index indicates state \\ 
& \quad $D$ -- donor $R$ -- receiver \\ 
$d^{R}(s)=d_{C}^{R}(s)=d_{N}^{R}(s)$ & mortality of passive receivers
depending on \\ 
& \quad the strategy of the donor \\ 
$C=d_{C}^{D}-d_{N}^{D}$ & cost of donor (depends on her action) \\ 
$B=d_{s}^{R}(N)-d_{s}^{R}(C)$ & benefit for receiver (depends on the action
of \\ 
& \quad the cooperative $(C)$ or noncooperative $(N)$ \\ 
& \quad donor) \\ 
$\tau ^{f}$ & intensity of the focal type of interaction \\ 
$x$ & number of receivers that can benefit from a single \\ 
& \quad cooperative action performed by the donor \\ 
$p_{s}^{R}$ & probability that a receiver with strategy $s$ \\ 
& \quad interacts with a cooperator \\ 
$p_{s}^{kin}$ & probability that a cooperative gene is carried \\ 
& \quad by the donor kin \\ 
$r$ & probability that receiver carries the same gene \\ 
& \quad from a common ancestor
\end{tabular}

\section{Introduction\protect\bigskip}

Kin selection and Hamilton's rule are described as among the most
important and influential concepts in modern evolutionary biology. These
concepts are popular in many disciplines where evolutionary reasoning is
used, such as evolutionary psychology. On the other hand, they are probably
the most misunderstood concepts in modern science \cite{Park 2007, West et al 2011}. In addition, the limits of their applicability are the subject of the
ongoing debate \cite{Fletcher et al 2006, Wenseleers 2006, Doebeli and Hauert 2006, West 2007, Van Veelen 2009}. After the release of a paper \cite{Nowak et al 2010}, this debate exploded with astonishing intensity
\cite{Rousset 2011, Gardner et al 2011, Allen 2015, Birch and Okasha 2015, Kramer and Meunier 2016, Okasha and Martens 2016a, Allen and Nowak 2016, Van Veelen et al 2017, Birch 2014, Birch 2017, Birch 2019, Bourke 2021, Koliofotis and Verreault-Julien 2022}.

The theory shows that for the spread of the altruistic gene, the famous
Hamilton's rule $Cost<Relatedness\ast Benefit$ should be satisfied. However
problem is that we have two main theoretical frameworks related to Hamilton's rule \cite{Van Veelen et al 2017}, one called the "counterfactual
method" \cite{Karlin and Matessi 1983, Matessi and Karlin 1984, Matessi and Karlin 1986} while
second is referred to as the "regression method" and it arises from the
Price equation \cite{Grafen 2006, Marshall 2015, Rousset 2015, Okasha 2016}. Those frameworks share similar terminology but define
them in a slightly different way. For example, relatedness in the
counterfactual method is defined as the probability that the cooperative
gene is inherited from the common ancestor, while in the regression method
as the regression coefficient. Also, there are other approaches where, for
example, the relatedness is defined as the probability that the receiver
carries the cooperative gene \cite{Nowak 2006} or ratio of probabilities \cite{Lehmann and Keller 2006}.

In addition, different alternative definitions of relatedness are mutually
compatible in so-called "additive payoff matrices" only (where differences
between row elements for both rows are equal and analogously for columns)
and require different cost and benefit definitions \cite{Van Veelen et al 2017, Van Veelen 2018}.\ This does not mean that these different approaches are
wrong. They probably have some limitations, but they are simply mutually
incompatible due to differences in their basic assumptions. The basic
underlying idea that individual cooperative behavior may support the spread
of the cooperative genes carried by other individuals is quite simple,
clear, and inspiring. However, the debate on this topic has become
increasingly complicated, and in turn, the mathematical formulations have
become very complex, as well, leading to the situation in which basic
questions about the meaning and sense of these concepts are still open
\cite{Marshall 2016}.

In this paper, we analyze relationships between Hamilton's rule and the
latest development in the evolutionary game theory related to demographic
games \cite{Argasinski and Broom 2013, Argasinski and Broom 2018a, Argasinski and Broom 2018b, Argasinski and Rudnicki 2021}. The difference between both frameworks is that in many problems
related to Hamilton's rule one interacting individual is active, and the
effects of his action are consumed by other passive actors. A similar
approach can be found in population genetics models \cite{Garay et al 2018a, Garay et al 2018b}. 

On the other hand, in the basic replicator dynamics framework, both players
exhibit their strategies and affect each other. This assumption is relaxed
in specific structured models on graphs, combining both perspectives \cite{Su et al 2022, Allen et al 2024}. Here, we will apply a similar approach to the
replicator dynamics models. Classical evolutionary games consist of a
game structure associated with replicator equations \cite{Maynard Smith 1982, Cressman 1992, Hofbauer and Sigmund 1988, Hofbauer and Sigmund 1998, Broom and Rychtar 2013, Friedman and Sinervo 2016, McNamara and Leimar 2020}. This approach is
mainly based on simple matrix games, where payoff matrices describe the
excess from the average growth rate in the population for the respective
strategies. To add the necessary ecological details and describe the models
in measurable parameters, the classical approach was expressed in terms of
demographic vital rates \cite{Argasinski and Broom 2013, Argasinski and Broom 2018a, Argasinski and Broom 2018b}. In this approach,
instead of a single payoff function, there are separate payoff functions
describing mortality and fertility. A similar explicit postulation of
opposed mortality and fertility forces as the cornerstone of the mechanistic
formulation (advocated by Geritz and Kisdi \cite{Geritz and Kisdi 2012}) of evolutionary theory was
proposed by Doebeli and Ispolatov \cite{Doebeli et al 2017}. 

However, this framework is not sufficient. The proposed approach is still
based on a very strong simplifying assumption. The individuals (and thus
their payoffs) differ only in terms of the inherited strategy, and
individuals carrying the same strategy are completely equivalent. Thus
births and deaths are not the only currency in which payoffs are paid in
evolutionary games.

An game theoretic approach, dealing with the problem of nonheritable
differences between individuals carrying the same strategies, was introduced
by Houston and McNamara \cite{Houston and McNamara 1999}. In the state-based approach, individual
differences caused by environmental conditions and their distribution in the
population are explicitly taken into consideration. Individuals and their
payoffs are determined by their actual state or situation (in our case this
is the donor/receiver role). This approach linking the replicator dynamics
with the state-based approach of Houston and McNamara was introduced in
\cite{Argasinski and Rudnicki 2021} by introduction of the state-switching
dynamics additional to the replicator equations. The special case of the
state-based models is the class of the age-structured evolutionary game
models \cite{Argasinski and Broom 2021}. The example of the state-switching
process was empirically observed among ants \emph{C.floridanus} \cite{Tripet and Nonacs 2004}. Older ants are more likely to forage, while younger
individuals are more focused on work within the nest. However, in this
paper, we will start from a much simpler problem of predator warning
signals, which will act as the basic illustrative example.

This paper is focused on the integration of this newly emerging synthetic
methodology with the very important concepts in the evolutionary theory of
Hamilton's rule and kin selection.\bigskip

\section{Goals of the paper}

The costs and benefits in Hamilton's rule can be expressed in
different ways, from changes in vital rates to changes in long-term
reproductive value. To be comparable with the demographic games, we need not
even the level of vital rates (which are the product of interaction rates
and demographic outcomes of interactions, \cite{Argasinski and Broom 2018a} but 
\textbf{the explicit demographic effect of the single interaction} (e.g., 
\textbf{change in survival probability}). Thus, we go to the extreme
opposition to the long-term reproductive value. Therefore, from the point of
view of a demographic event-based approach \cite{Argasinski and Broom 2013, Argasinski and Broom 2018a, Argasinski and Broom 2018b}, most of the considered cases are probably related to some
type of danger (such as helping a drowning individual) or to energy gain and
expenditure that also affect current survival (such as altruism among
vampire bats) and have no direct reproductive output. However, exist cases
linked with reproductive success (for example, plants attracting
pollinators; \cite{Sun et al 2021}). Therefore, the goals of the paper are as
follows.\bigskip 

\subsection{Preliminary results: initial "null" alarm call model and the
issue of non-additive problems}

In the preliminary technical results, we describe the existing classic
Hamilton's rule and kin selection frameworks in demographic parameters to
make them compatible with the demographic game approach and, later,
comparable with the main result of the paper. Careful derivation is
necessary because it is not obvious that the interaction rates and
background vital rates will cancel out in the resulting rule for
cooperation. In addition, while using standard theory, this will be a
helpful example for readers unfamiliar with those frameworks.
We consider the predator alarm call as the illustrative conceptual example
\cite{Maynard Smith 1965, Tamachi 1987, Taylor et al 1990, de Assis et al 2018}. This is the classic example of altruistic behavior, supported by
empirical observations \cite{Dunford 1977, Sherman 1977, Hoogland 1983, Griesser 2013}. Therefore, in our model, we use mortality payoffs to describe the
costs of the sacrifice of the donor individual and the benefits resulting
from the rescue of the receiver. The obtained model will be used for the
analysis of the widely discussed issues related to the non-additivity of
payoff functions. \bigskip

\subsection{Main result: model with explicit fluxes between donor/receiver
roles and the rule for cooperation under state distribution asymmetry}

Classical theory (and our preliminary "null" model) contains the silent
assumption, that the distribution of roles is constant and the
donor/receiver role is randomly drawn during each interaction event. This
assumption is certainly satisfied in many cases, however we can imagine
situations when it is not applicable. We will extend the initial "null"
model to the case when the distribution of states is not constant but is the
product of some dynamic processes (fluxes between receiver and donor roles),
which is described by additional equations \cite{Argasinski and Rudnicki 2021}.
Then, we derive the general rule for cooperation from the extended dynamic
model, which will constitute the main result of the paper.\bigskip

\subsection{Methods}

We will combine the Hamilton's methodology with the demographic
approach to Evolutionary Games \cite{Argasinski and Broom 2013, Argasinski and Broom 2018a, Argasinski and Broom 2018b} and
State Switching Dynamics for game theoretic models based on state
\cite{Argasinski and Rudnicki 2021}. In this paper, we adopt a counterfactual
method for derivation of the Hamilton's rule models \cite{Karlin and Matessi 1983, Matessi and Karlin 1984, Matessi and Karlin 1986} since this methodology can use the same
parameters as a demographic games. The necessary details related to
Hamilton's rule are in Appendix 1, discussion of possible misunderstandings
of it are in Appendix 2 and the necessary basic details of the demographic
game approach are in Appendix 3.\bigskip 

\section{Part one, preliminary results: "null" model based on the classical
theory described in terms of demographic games\protect\bigskip}

Here, we derive the "null" model mentioned in section 2.1, based on the
standard theory, which will be later generalized to develop the main result.
This will provide a platform for comparison of the new results with
classical theory from the literature. To achieve this goal, we derive a
demographic equivalent of the "donation game" \cite{Marshall 2015}, a model of
altruistic sacrifice expressed in terms of the average mortality changes
during the focal event. We will focus on the distribution of the receiver
and donor roles among individuals. In addition, we will discuss a payoff
non-additivity issue. Let us consider the problem of signaling the predator
threat (emission of the signal may help others while exposing the signalist
to the increased danger) as the conceptual example for our framework. Assume
that a random member of the population may spot the attacking flying
predator and warn neighbors (see Figure 1).\bigskip

\begin{figure}[]
\centering
\includegraphics[width=9cm]{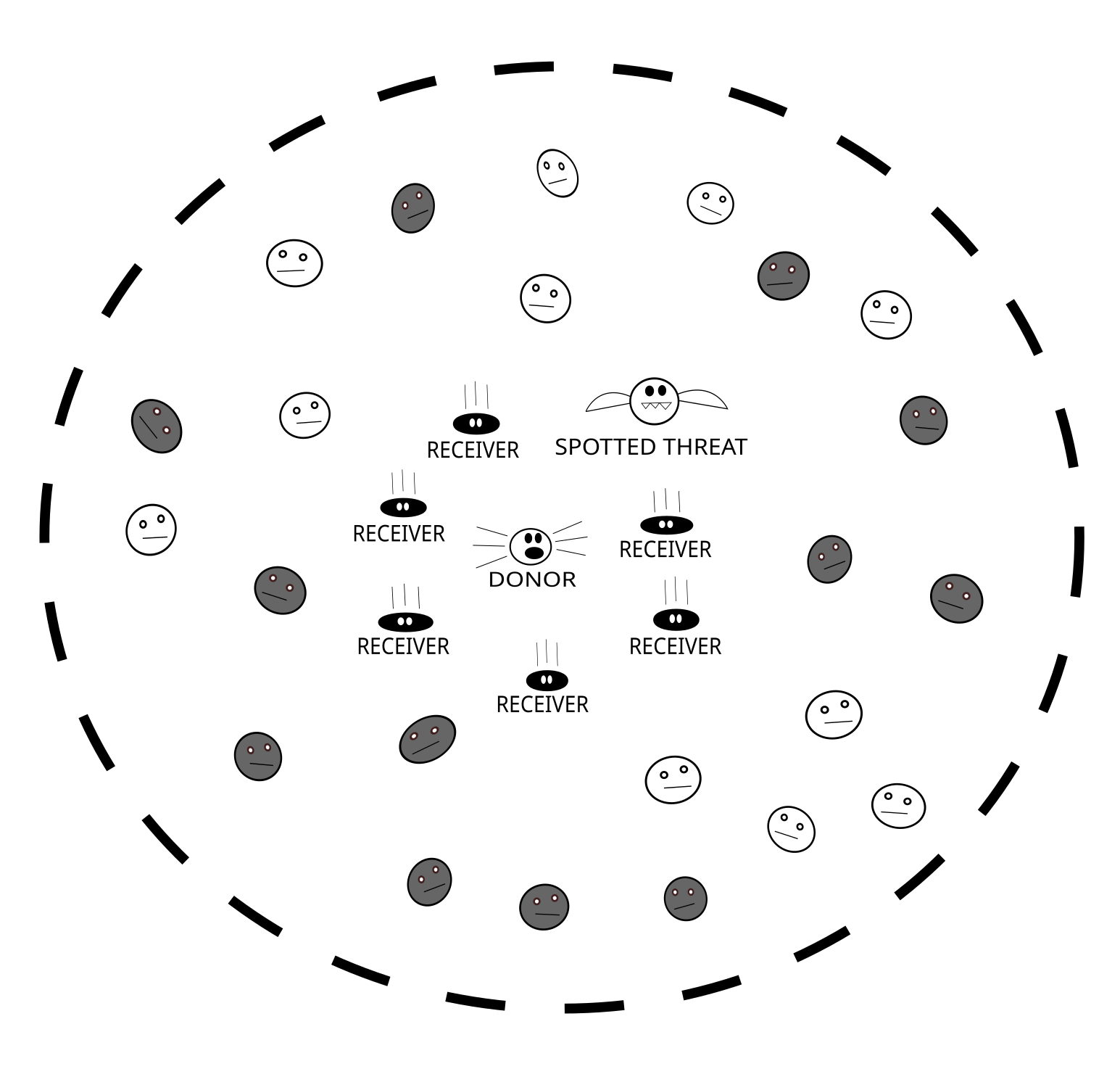}
\caption{ \emph{Warning signal example (cooperators are white and
non-cooperators grey): cooperating donor who spotted the threat may warn
assorted receivers. Emission of the warning signal may expose the
cooperator, leading to increased mortality, constituting cost.}}
\end{figure}

Then, we have trade-offs between the donor's mortality and the receivers'
expected survival. In this case, the background growth rate $R_{b}$ should
be the same in both states and will contain all fertility effects. We have
two roles or states of individuals (passive donor and active receiver) and
two competing strategies (cooperative and noncooperative). Strategies are
heritable, while the assigment of the donor/receiver role is random,
non-heritable, and independently drawn at each interaction.\ Only active
donors reveal their strategies (helping or not helping receivers). Receivers
are passive; thus, their strategy is latent and cannot be observed during
the focal interaction. Therefore, the logical outcome of these assumptions
is that each altruistic act should be associated with the same cost for the
donor and the resulting benefit for the receiver.

The receiver's mortality payoff is independent of the carried strategy,
while donors can exhibit two types of behavior: cooperate (pay the
cost), described by subscript $C$, or defect (do not pay the cost), described by subscript $N$.\ Then, we have the payoff functions $
d_{s}^{R}(a)$, where subscript $s$ describes the carried strategy,
superscript $R$ describes the donor/receiver role and argument $a$ describes
the action of the individual acting in the opposite role. The donor's
mortality $d_{N}^{D}$ and $d_{C}^{D}$ depends on her action only (\textbf{\
thus argument} $(a)$ \textbf{is an obsolete element of notation in her
payoffs}). Then the cost can be expressed as 
\begin{equation}
C=d_{C}^{D}-d_{N}^{D}>0,\text{ \ \ thus \ \ }d_{C}^{D}=d_{N}^{D}+C.
\label{payoff}
\end{equation}
Then, since the receivers' mortality depends only on the action of the
possible donors, we have

\begin{equation*}
d_{C}^{R}(a)=d_{N}^{R}(a)=d^{R}(a)
\end{equation*}
(\textbf{thus, in this case, the strategy subscript is an obsolete element
of notation}). Since receivers are passive, their payoff functions are the
same for both strategies. A single receiver of the cooperative behavior has
payoff $d^{R}(C)$ in comparison to receivers not affected by the cooperative
behavior, which will have mortality $d^{R}(N)$. Since $d^{R}(N)>d^{R}(C)$,
we can define the benefit of the receiver as 
\begin{equation}
B=d^{R}(N)-d^{R}(C)>0
\end{equation}
(leading to $d^{R}(C)=d^{R}(N)-B$ since the benefit describes a decrease in
mortality). In many cases, such as giving a predator warning signal, a
single cooperative donor can alarm a few receivers with different
strategies. Let us begin with the assumption of well-mixed population and
frequency-dependent selection, as in evolutionary games,\ to see the
limitations of this approach in this case. Assume that $x$ is the number of
receivers that can be affected by the behavior of the single donor. Then,
for both strategies, we have the same constant distribution of states (the
exceptions to this assumption are the subject of the second part of this
paper), described by $q^{D}=1/\left( 1+x\right) $ (while $q^{R}=1-q^{D}$ is
the fraction of receivers). Donor/receiver role is randomly drawn each time
the threat occurs.\ Note that only in the case of donors do we have
different payoffs for different strategies.\ The threat, such as hunting
predator, is interpreted as the focal event. Then randomly chosen individual
can spot the approaching predator and according to the carried strategy it
can warn neighboring individuals or not. Therefore, cooperating donor is
drawn with probability proportional to the fraction of cooperators in the
population $g_{C}$. Then the receiver's average mortality\ payoff will be 
\begin{eqnarray}
d^{R} &=&g_{C}\left( d^{R}(N)-B\right) +\left( 1-g_{C}\right) d^{R}(N) 
\notag \\
&=&d^{R}(N)-g_{C}B,  \label{mortR}
\end{eqnarray}

the same for both strategies, since receivers are passive. Therefore,
according to (\ref{payoff}) the only difference in mortality payoffs between
cooperators and noncooperators is caused by cost of altruism$\ C$ \. This
shows that in the random matching of the individuals (as in the classical
game theory), cooperators have greater mortality than noncooperators.
Therefore, a cooperative strategy will lose. Thus cooperators
should not help everyone, but should support other cooperators. Let us relax
the assumption of a panmictic population and add the assortment mechanism
and later kin selection to our model. Assortment mechanisms are primarily
based on kinship and family structures or sometimes on other types of
structured populations \cite{Cavalli-Sforza and Feldman 1978, Grafen 1979, Eshel and Cavalli-Sforza 1982, Allen and Nowak 2015}. In this paper we limit
ourselves to the basic kin selection case. Otherwise, we will not specify
the assortment mechanism. Thus, a cooperating receiver receives help with
probability $p_{C}^{R}$, while a noncooperative receiver receives help with
probability $p_{N}^{R}$. Let us update the mortality functions. In this
case, analogous to (\ref{mortR}), the impact on the receiver is different for
both strategies:

\begin{align}
d_{C}^{R}={}& p_{C}^{R}\left( d^{R}(N)-B\right) +\left( 1-p_{C}^{R}\right)
d^{R}(N)  \notag \\
& =d^{R}(N)-p_{C}^{R}B,  \label{mortC}
\end{align}
\begin{align}
d_{N}^{R}={}& p_{N}^{R}\left( d^{R}(N)-B\right) +\left( 1-p_{N}^{R}\right)
d^{R}(N)  \notag \\
& =d^{R}(N)-p_{N}^{R}B.  \label{mortN}
\end{align}
Then the average payoff of $s$-strategist is 
\begin{equation}
d_{s}^{f}=q^{D}d_{s}^{D}+\left( 1-q^{D}\right) d_{s}^{R}(p_{s}^{R}),
\end{equation}
and for both competing strategies the above functions are 
\begin{eqnarray}
d_{C}^{f} &=&q^{D}\left[ d_{N}^{D}+C\right] +\left( 1-q^{D}\right) \left[
d^{R}(N)-p_{C}^{R}B\right]  \label{focalC} \\
&=&q^{D}d_{N}^{D}+\left( 1-q^{D}\right) d^{R}(N)-\left( 1-q^{D}\right)
p_{C}^{R}B+q^{D}C  \notag \\
d_{N}^{f} &=&q^{D}d_{N}^{D}+\left( 1-q^{D}\right) \left[ d^{R}(N)-p_{N}^{R}B 
\right]  \label{focalN} \\
&=&q^{D}d_{N}^{D}+\left( 1-q^{D}\right) d^{R}(N)-\left( 1-q^{D}\right)
p_{N}^{R}B.  \notag
\end{eqnarray}

Functions $d_{C}^{f}$ and $d_{N}^{f}$ differ only by terms $-\left(
1-q^{D}\right) p_{C}^{R}B+q^{D}C$ and $-\left( 1-q^{D}\right) p_{N}^{R}B$.
We can present our framework in the matrix game form (derivation in Appendix
4)\ where entries describe "fitness effects" (differences in payoffs
resulting from the particular action):\bigskip

\begin{equation*}
\left[ 
\begin{array}{cc}
0 & -\left( 1-q^{D}\right) B \\ 
q^{D}C & q^{D}C-\left( 1-q^{D}\right) B
\end{array}
\right] .
\end{equation*}

For $q^{D}=0.5$, we have a donation game matrix \cite{Panchanathan and Boyd 2003, Marshall 2015} 
\begin{equation}
0.5\left[ 
\begin{array}{cc}
0 & -B \\ 
C & C-B
\end{array}
\right] .
\end{equation}

which in the classical game theoretical approach should be multiplied by a
vector of strategy frequencies $[1-g_{c},g_{c}]^{T}$. However, when
assumption of well mixed population is relaxed, frequencies should be
replaced by elementwise multiplication by a matrix of assortment
probabilities (defined in Appendix 1): 
\begin{equation*}
\left[ 
\begin{array}{cc}
1-p_{N}^{R} & p_{N}^{R} \\ 
1-p_{C}^{R} & p_{C}^{R}
\end{array}
\right] 
\end{equation*}

leading to row payoffs $-p_{N}^{R}B$ and $C-p_{C}^{R}B$. The structure of
this matrix results from the underlying assumptions that both fitness
effects are products of a single altruistic act performed by a cooperative
donor. The second assumption is that receivers are passive, and their
strategies are latent and thus indistinguishable. Every "synergistic effect"
between a cooperative donor and a cooperative receiver violates the second
assumption. Therefore, for this class of problems, we don't need
"nonadditive" models, and this is not a limitation of the counterfactual
method (see discussion in Appendix 4).

Assume that the focal event occurs at the intensity $\tau _{f}=1$. Then we
can formulate the growth equations (\ref{basic2}) from Appendix 3: 
\begin{align}
\dot{n}_{C}& =n_{C}\left( R_{b}-d_{C}^{f}\right)  \\
\dot{n}_{N}& =n_{N}\left( R_{b}-d_{N}^{f}\right) ,
\end{align}
For greater growth rate of cooperators we need $\left(
R_{b}-d_{C}^{f}\right) >\left( R_{b}-d_{N}^{f}\right) $. Then $R_{b}$ and $
q^{D}d_{N}^{D}+\left( 1-q^{D}\right) d^{R}(N)$ from $d_{C}^{f}$ (\ref{focalC}) and $d_{N}^{f}$\ (\ref{focalN}) will cancel out. This leads to 
\begin{equation*}
\left( 1-q^{D}\right) p_{C}^{R}B-q^{D}C>\left( 1-q^{D}\right) p_{N}^{R}B,
\end{equation*}
leading to the classical condition 
\begin{equation}
\left[ p_{C}^{R}-p_{N}^{R}\right] \frac{\left( 1-q^{D}\right) }{q^{D}}B>C
\label{StrongHam}
\end{equation}
which is also known from the literature version of equation (\ref{StrongHam}) for multiple receivers (and $\left( 1-q^{D}\right) /q^{D}$ describes the
number of receivers per single donor).\emph{\ }

For the kin selection case $p_{C}^{R}$ and $p_{N}^{R}$ are replaced by
probabilities $p_{C}^{kin}$ and $p_{N}^{kin}$ that kin donor inherited the
cooperative gene (see Appendix 1).\ Since $p_{C}^{kin}-p_{N}^{kin}=r$ where $
r$ is relatedness, formula (\ref{StrongHam}) becomes 
\begin{equation}
r\frac{\left( 1-q^{D}\right) }{q^{D}}B>C,  \label{HamKin}
\end{equation}

 which is the classical Hamilton's rule.

Interaction rates and background growth rates cancel out and do not
affect the fitness effects describing cost and benefit. Then, cost $C$ 
 and benefit $B$ are expressed in terms of the focal
interaction's average mortality instead of abstract fitness, reproductive
value, or even vital rates. In our model, reproduction is realized by the
background growth rate, and there is no need to take it into account.
Therefore, we have a model based on the classical theory described in terms
compatible with the demographic game approach. This leads to the interesting
property. Note that the application of the approach, where fitness effects
are expressed in the "number of offspring equivalents" (as it is defined in
the Encyclopedia Britannica) or reproductive value (as, for example, in
\cite{Marshall 2015}), technically implies the calculus of unborn offspring. This
is hardly testable. When we reduce the generality of the model by replacing
the general fitness parameter with a specific demographic payoff, such as
mortality (as in our model), the obtained framework can act as the
predictive model, not only as the abstract theorem. It seems that it is
necessary to derive real-life, falsifiable models. The resulting model can
be parameterized by simple statistical mortality estimation based on
observations of the focal interactions (for example, see \cite{Griesser 2013}).

We use the alarm call problem as an illustrative example. However, the
obtained formalism can be used for all problems where behavioral traits
determine survival only. When necessary, the survival payoff can be replaced
by fertility or the more complicated trade-offs between them used in the
demographic game-theoretic models \cite{Argasinski and Broom 2013, Argasinski and Broom 2018a,
Argasinski and Broom 2018b}. The number of receivers from the classical theory is equivalent to
the donor/receiver role distribution from the state-based evolutionary game
\cite{Argasinski and Rudnicki 2021}, described by parameter $q^{D}$. In the
classical theory, $q^{D}$ is constant, and the role is independently drawn
at each focal interaction. In the next section, we build the model based on
the state-switching dynamics \cite{Argasinski and Rudnicki 2021}, where this
assumption is relaxed. Then the role distribution is the product of some
external population process. \bigskip

\section{Part two, main results: explicit dynamics of donor/receiver roles}

\subsection{Rationale for part two}

Note that the analysis of the problem of altruism was limited to the simple
system of exponential growth equations. In the previous sections, the
distribution of roles was determined by the conditional probability of
acting as a donor or receiver related to the focal interaction. This should
be correct in many cases when the role is strictly limited to the particular
game round and in the next round is independently drawn again. However, it
is also possible that the donor or receiver role is determined by some
external conditions and in consequence cannot be changed in the focal
interaction. For example, a vampire bat foraging in areas where the
abundance of prey is very low needs support until it finds an area where
prey abundance is high, which it may exploit for some time. Altruistic
behavior may increase the survival of the receiver, but it cannot help it
find the source of food. Similarly, in the case of a predator warning signal
(our main conceptual example), we can imagine that the population is
structured and divided into two groups, one of which is\ more exposed to the
observation of the threat (for example, due to being at the border of the
habitat). However, the exposed individuals, according to their strategy, can
warn other individuals or not and after the warning event can move to
another location or stay at the border of the habitat. The mobility may be
completely independent of the results of the focal event. This division may
not be fixed, and the individuals may randomly shift between different roles
(see Figure 2).

\begin{figure}[]
\centering
\includegraphics[width=9cm]{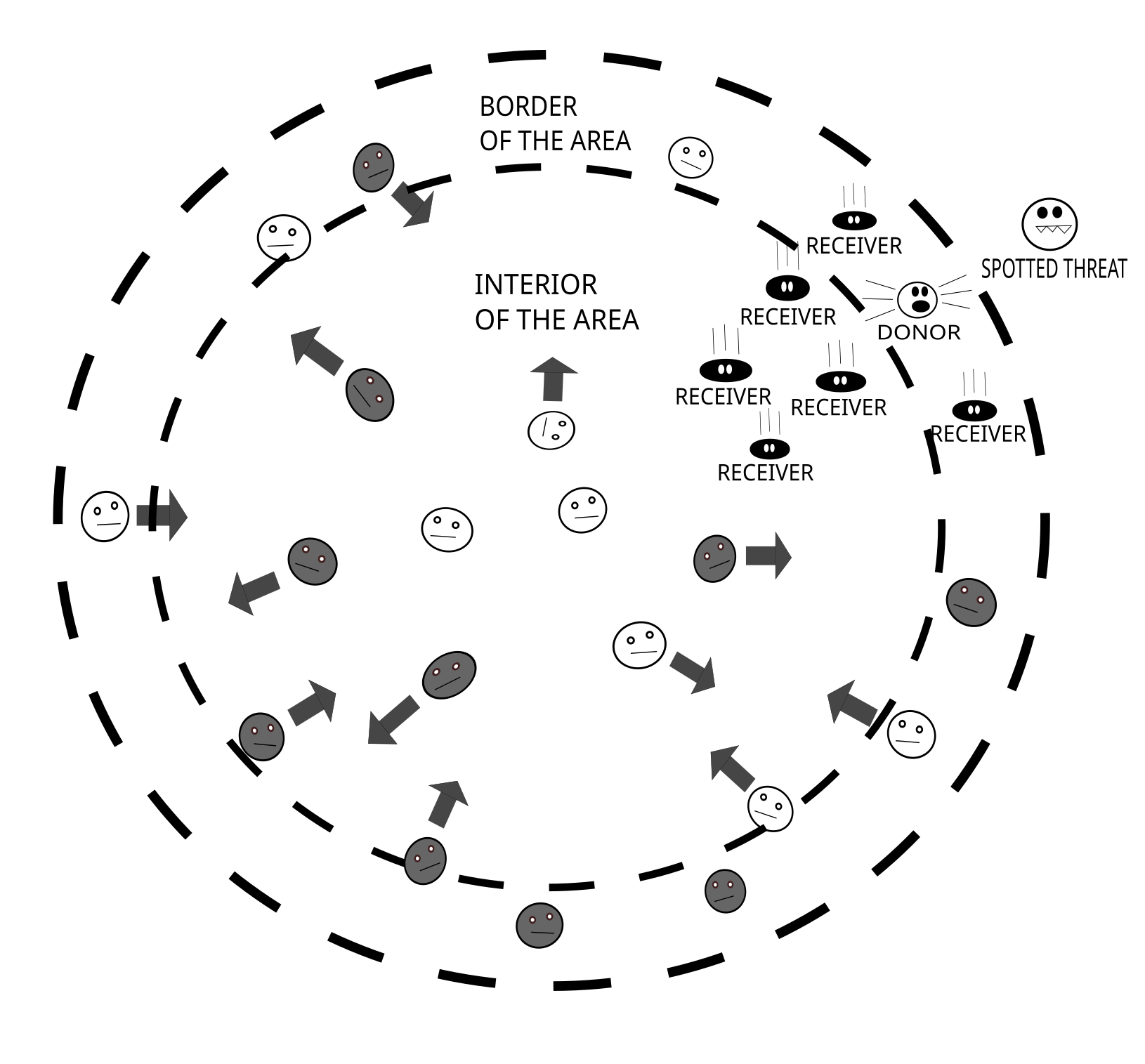}
\caption{ \emph{The case when only individuals from the border of the
habitat may spot the threat and warn the assorted receivers (cooperators are
white and non-cooperators grey). Individuals move between habitat interior
and border according to some mechanistic rules and role is not randomly
chosen each time but is determined by the current positions of the
individuals. Surviving cooperator may warn neighbors again until she moves
toward the interior and becomes the potential receiver.}}
\end{figure}

This leads to a separate population process and background switching
dynamics that may depend on the daily movement routines of individuals.
Therefore, we can imagine that the population structure (distribution of
roles) emerges as a dynamic equilibrium of some independent process. This
process is driven by some basic principles describing the fluxes of
individuals between those roles. Phenomena of this type can be termed \textbf{dynamically structured populations}. Then we can use our framework
to extend the static reasoning to the dynamic case where the distribution of
roles varies in time. In this case, we should describe the respective
dynamics for both strategies and the evolution of the distribution of states
for each strategy.

\subsection{Derivation of the replicator dynamics with explicit dynamics of
fluxes between states}

In this section, we use the role-switching dynamics (across two opposite
states) introduced in \cite{Argasinski and Rudnicki 2021}. Below we re-derive the
simplest case of this approach. Our opposing states are donor and receiver.
We extend our dynamics by means of explicit background intensities (i.e. not
related to the threat and the altruistic act) of switching between roles,
described by $\Lambda ^{i}$ as the background intensity of leaving role $i$
and taking on the opposite role. Note that the parameters $\Lambda ^{i}$ may
be not constants but functions of the actual distribution of roles in the
population, described by $g_{1}q_{1}^{1}+(1-g_{1})q_{2}^{1}$ ; however, for
simplicity, we do not describe this explicitly in the formalism. For
simplicity, assume that $R_{b}+R_{s}^{1}$ describes the overall Malthusian
growth rate (sum of the density-dependent background fitness and focal game
payoffs) for strategy $s$ acting in role $1$. Then, the growth equation for
strategy $s$ in role $1$ can be described as:

\begin{equation}
\dot{n}_{s}^{1}=n_{s}^{1}\left[ R_{b}+R_{s}^{1}-\Lambda ^{1}\right]
+n_{s}^{2}\Lambda ^{2}  \label{malth1}
\end{equation}
and\ the analogous equation for role 2. We use the multi-population approach
to replicator dynamics \cite{Argasinski 2006, Argasinski 2012, Argasinski 2013}, where the population
can be decomposed into subpopulations described by their own replicator
dynamics. Then subsystems describing the subpopulations are completed
through an additional set of replicator equations describing the dynamics of
the proportions of all subpopulations. Then, we can describe the
distribution of states among $s$-strategists in related frequencies $
q_{s}^{i}=n_{s}^{i}/\sum_{j}n_{s}^{j}$. In the special case in which for all
strategies, we have only two states, we can apply the well-known form of
replicator dynamics for two strategies, albeit applied in our case not for
strategies but for separate roles among carriers of some strategy (described
by upper superscript). Then the above system reduces to a single equation
(detailed derivation in Appendix 5):

\begin{equation}
\dot{q}_{s}^{1}=q_{s}^{1}(1-q_{s}^{1})\left[ R_{s}^{1}-R_{s}^{2}\right] + 
\left[ \left( 1-q_{s}^{1}\right) \Lambda ^{2}-q_{s}^{1}\Lambda ^{1}\right] .
\label{2strategy}
\end{equation}
Now, we can describe the selection of strategies through the application of
the multipopulation replicator dynamics. Then, the above system should be
completed by the additional set of replicator equations describing the
relative frequencies of the other strategies. As was shown in \cite{Argasinski and Rudnicki 2021}, the dynamics of state changes do not have a direct impact
on the strategy frequencies (or on the population size) since they do not
change the number of strategy carriers ($\Lambda $\ terms cancel out in
equations for strategy frequencies). Then, we have the following system
describing the selection:

\begin{equation}
\dot{g}_{1}={}g_{1}(1-g_{1})\left[ \bar{R}_{1}(q_{1})-\bar{R}_{2}(q_{2}) 
\right] ,  \label{selectionEQ}
\end{equation}
where 
\begin{equation}
\bar{R}_{s}(q_{s})=q_{s}^{1}R_{s}^{1}+(1-q_{s}^{1})R_{s}^{2}.
\end{equation}
The above system should be completed by the equation on the total population
size (the only element where background growth rate $R_{b}$ is present): 
\begin{equation}
\dot{n}=n\left[ R_{b}+g_{1}\bar{R}_{1}(q_{1})+(1-g_{1})\bar{R}_{2}(q_{2}) 
\right] ,  \label{popsize}
\end{equation}

and explicit density dependence is omitted here for simplicity.\bigskip

\subsection{The dynamics of altruism}

Now, we can update our model from the first part of the paper to the case
describing the dynamics of roles. For the description of the rules
underlying the state changes, we can use the background switching dynamics.
Background intensities of leaving donor and receiver roles are $\Lambda ^{D}$
\ and $\Lambda ^{R}$ respectively. The switching term (\ref{switchingterm})
describing the fluxes between donor and receiver roles has the form $\left( 
\dfrac{n_{s}^{R}}{n_{s}^{D}}\Lambda ^{R}-\Lambda ^{D}\right) $. Recall that
we assumed that the focal interaction happens at intensity $\tau ^{f}$,
which was removed by changing the timescale. We assumed that the switching
of roles is independent of the results of the focal interaction. When $
\Lambda ^{D}<\tau ^{f}$ and $\Lambda ^{R}<\tau ^{f}$ the number of role
switches is smaller than the number of rounds in that game (\textbf{maximum
one switch per focal game round/interaction}) This leads to $\Lambda ^{D}<1$
and $\Lambda ^{R}<1$ after the change in the timescale. Then the cooperative
donor may warn receivers multiple times before role switch. For $\Lambda
^{D}>1$ and $\Lambda ^{R}>1$ role circulation is faster than threat
occurrence and cooperator sometimes may be exposed to the threat, while most
donor periods are safe. Therefore, growth equations (\ref{malth1}) for the
competing strategies will have form: 
\begin{eqnarray*}
\dot{n}_{s}^{D} &=&n_{s}^{D}\left( R_{b}-d_{s}^{D}+\left( \dfrac{n_{s}^{R}}{
n_{s}^{D}}\Lambda ^{R}-\Lambda ^{D}\right) \right) \\
\dot{n}_{s}^{R} &=&n_{s}^{R}\left( R_{b}-d_{s}^{R}+\left( \dfrac{n_{s}^{D}}{
n_{s}^{R}}\Lambda ^{D}-\Lambda ^{R}\right) \right) .
\end{eqnarray*}

Above equations transformed into the state switching dynamics (\ref
{2strategy}) constitute the following subsystem (derivation in Appendix 6)

\begin{align}
\dot{q}_{C}^{D}={}& \left( \left( 1-q_{C}^{D}\right) \Lambda
^{R}-q_{C}^{D}\Lambda ^{D}\right)  \notag \\
& {}-q_{C}^{D}\left( 1-q_{C}^{D}\right) \left[ d_{N}^{D}+C-\left(
d^{R}(N)-p_{C}^{R}B\right) \right] ,  \label{q-C}
\end{align}
\begin{align}
\dot{q}_{N}^{D}={}& \left( \left( 1-q_{N}^{D}\right) \Lambda
^{R}-q_{N}^{D}\Lambda ^{D}\right)  \notag \\
{}& -q_{N}^{D}\left( 1-q_{N}^{D}\right) \left[ d_{N}^{D}-\left(
d^{R}(N)-p_{N}^{R}B\right) \right] .  \label{q-N}
\end{align}
It is clear that these dynamics lead to different role distributions for
different strategies. How does this affect the selection process? Let us
derive the replicator dynamics describing the selection of the strategies.
This leads to the average mortalities (see Appendix 7 for the derivation): 
\begin{eqnarray}
d_{N}^{f} &=&q_{N}^{D}d_{N}^{D}+\left( 1-q_{N}^{D}\right) d_{N}^{R}  \notag
\\
&=&q_{N}^{D}d_{N}^{D}+\left( 1-q_{N}^{D}\right) \left(
d^{R}(N)-p_{N}^{R}B\right)  \label{averN} \\
d_{C}^{f} &=&q_{C}^{D}d_{C}^{D}+\left( 1-q_{C}^{D}\right) d_{C}^{R}  \notag
\\
&=&q_{C}^{D}\left( d_{N}^{D}+C\right) +\left( 1-q_{C}^{D}\right) \left(
d^{R}(N)-p_{C}^{R}B\right) ,  \label{averC}
\end{eqnarray}

and the resulting selection equation 
\begin{align}
\dot{g}_{C}& =g_{C}\left( 1-g_{C}\right) \left(
d_{N}^{f}(q_{N}^{D})-d_{C}^{f}(q_{N}^{D})\right)  \notag \\
& =g_{C}\left( 1-g_{C}\right) \left[ \left( q_{N}^{D}-q_{C}^{D}\right)
\left( d_{N}^{D}-d^{R}(N)\right) \right.  \notag \\
& \left. +\left[ \left( 1-q_{C}^{D}\right) p_{C}^{R}-\left(
1-q_{N}^{D}\right) p_{N}^{R}\right] B-q_{C}^{D}C\right] .  \label{genedyn}
\end{align}

which together with switching dynamics (\ref{q-C}) and (\ref{q-N}) will
constitute our general modeling framework. If it is necessary, the above
system can be completed by an additional equation describing the dynamics of
the population size.

\section{When cooperative strategy wins?}

\begin{figure}[!h]
\centering
\includegraphics[width=11cm]{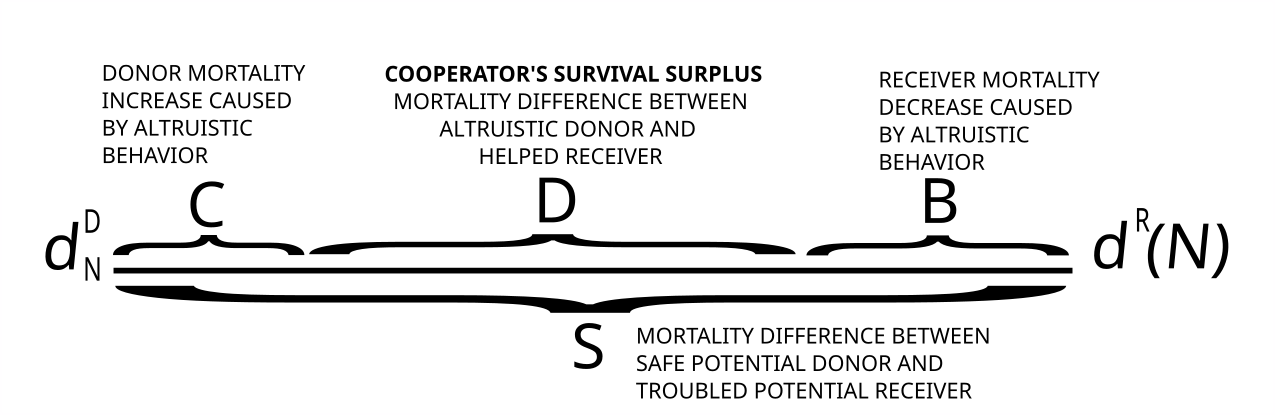}
\caption{ \emph{Diagram showing fitness effects resulting from mortality
differences. $S=d^{R}(N)-d_{N}^{D}$ and $D=d^{R}(C)-d_{C}^{D}$, in effect $
S=B+D+C$.}}
\end{figure}
\bigskip

We can express the mortality functions (\ref{averN}) and (\ref{averC})\ in
terms of the differences in mortalities constituting the fitness effects
used in the cost vs benefit calculus. It is reasonable to assume that\ the
mortality of the noncooperator in the role of the donor should be equal to
or smaller than the mortality of the receiver receiving help. Doing nothing
cannot be more dangerous than being rescued. Then, we have that $
d_{N}^{D}<d^{R}(C)$. Thus, we should interpret the factor $
S=d^{R}(N)-d_{N}^{D}$. This leads to 
\begin{equation}
d_{N}^{D}\leq d^{R}(C)=d^{R}(N)-B,
\end{equation}
leading to 
\begin{equation}
B\leq S=d^{R}(N)-d_{N}^{D}>0,
\end{equation}
Therefore, the parameter $S$ is simply a benefit of not being in trouble
(which means being in the role of receiver). Since $d_{N}^{D}=d_{C}^{D}-C$,
parameter $S$ can be presented as $S=B+D+C$ where $D=d_{C}^{D}-d^{R}(C)$
(see Figure 3 for the meaning of these parameters). Then, 
\begin{equation*}
d^{R}(N)=d_{N}^{D}+B+D+C
\end{equation*}
and thus $D$ describes the difference in mortalities between the helping
donor and the helped receiver. Then the mortality functions (\ref{averN})
and (\ref{averC}) will be (derivation in Appendix 8a):

\begin{eqnarray*}
d_{C}^{f} &=&q_{C}^{D}\left[ d_{N}^{D}+C\right] +\left( 1-q_{C}^{D}\right) 
\left[ d^{R}(N)-p_{C}^{R}B\right] \\
&=&d_{N}^{D}+\left( 1-q_{C}^{D}\right) \left( \left( 1-p_{C}^{R}\right)
B+D\right) +C
\end{eqnarray*}
\begin{eqnarray*}
d_{N}^{f} &=&q_{N}^{D}d_{N}^{D}+\left( 1-q_{N}^{D}\right) \left[
d^{R}(N)-p_{N}^{R}B\right] \\
&=&d_{N}^{D}+\left( 1-q_{N}^{D}\right) \left( \left( 1-p_{N}^{R}\right)
B+D+C\right)
\end{eqnarray*}

Structure of those functions differ by factor the $q_{C}^{D}C$, describing
the cost payed by the cooperative donors. From the condition $
d_{N}^{f}(g,q)-d_{C}^{f}(g,q)>0$, resulting from (\ref{selectionEQ}), we can
derive the rule for the increase in cooperation describing the relationships
between cost and benefit. From the bracketed term from (\ref{genedyn}), we
have that it is

\begin{equation}
\left( 1-q_{N}^{D}\right) \left( \left( 1-p_{N}^{R}\right) B+D+C\right)
>\left( 1-q_{C}^{D}\right) \left( \left( 1-p_{C}^{R}\right) B+D\right) +C
\label{main}
\end{equation}

\clearpage

\subsection{\textbf{Main result:} the general rule for cooperation under
state distribution asymmetry\protect\bigskip}

Equation (\ref{main}), leads to the\textbf{\ general rule for cooperation}
(derivation in Appendix 8b) expressed in terms of $B$, $C$ and $D$ (meaning
of those parameters is explained in Figure 3): 
\begin{equation}
\left[ \left( 1-q_{N}^{D}\right) \left( 1-p_{N}^{R}\right) -\left(
1-q_{C}^{D}\right) \left( 1-p_{C}^{R}\right) \right] B+\left[
q_{C}^{D}-q_{N}^{D}\right] D>q_{N}^{D}C  \label{coop2}
\end{equation}

and factors $\left( 1-q_{N}^{D}\right) \left( 1-p_{N}^{R}\right) $ 
and $\left( 1-q_{C}^{D}\right) \left(1-p_{C}^{R}\right) $ describe the fractions of unhelped
individuals of both strategies. \bigskip

Therefore, $D>0$, in addition to the cost $C$ saved by noncooperator, can be
termed the \textbf{cooperator's survival surplus}. This may happen when the
cooperator can secure the safe shelter before the emission of the warning
signal, which leads to a survival advantage over assorted neighbors. Note
that for equal role distributions for both strategies (thus $
q_{N}^{D}=q_{C}^{D}=q^{D}$), equation (\ref{coop2}) reduces to the classical
Hamilton's rule (\ref{StrongHam}):

\begin{equation}
\left[ p_{C}^{R}-p_{N}^{R}\right] \frac{\left( 1-q^{D}\right) }{q^{D}}B>C
\end{equation}

For simplicity we can assume that $d_{N}^{D}=0$ and $d^{R}(N)=1$, leading to 
$D=1-B-C$. Then the average mortalities (\ref{averC}) and (\ref{averN}) can
be presented in the form of the \emph{relative fitness effect surfaces }
describing the normalized differences in mortality (derived in Appendix 8c):

\begin{equation}
d_{C}^{f}=\left( 1-q_{C}^{D}\right) \left( 1-p_{C}^{R}B\right) +q_{C}^{D}C
\label{relfitC}
\end{equation}

\begin{equation}
d_{N}^{f}=\left( 1-q_{N}^{D}\right) \left( 1-p_{N}^{R}B\right) .
\label{relfitN}
\end{equation}

Since the rule for cooperation depends on the role distributions, we can
assume the simplest example when switching rates $\Lambda ^{R}$\ and $
\Lambda ^{D}$\ are constant. For nonzero paraameters the state switching
dynamics has single stable restpoint in the interior of $(0,1)$ and has the
form  (derivation in Appendix 9a): 
\begin{eqnarray*}
\dot{q}_{C}^{D} &=&\left( \left( 1-q_{C}^{D}\right) \Lambda
^{R}-q_{C}^{D}\Lambda ^{D}\right) +q_{C}^{D}\left( 1-q_{C}^{D}\right) \left[
\left( 1-p_{C}^{R}\right) B+D\right] \\
\dot{q}_{N}^{D} &=&\left( \left( 1-q_{N}^{D}\right) \Lambda
^{R}-q_{N}^{D}\Lambda ^{D}\right) +q_{N}^{D}\left( 1-q_{N}^{D}\right) \left[
\left( 1-p_{C}^{R}\right) B+C+D\right] .
\end{eqnarray*}

Then we can calculate the rest points of the switching dynamics (\ref{q-C})
and (\ref{q-N}), constituting the stable role distributions (Appendix 9a). Those unique stable restpoints have the following general
form: 
\begin{eqnarray*}
\tilde{q}_{s}^{D} &=&\frac{-\left[ \Lambda ^{R}+\Lambda ^{D}-A_{s}\right] + 
\sqrt{\left[ \Lambda ^{R}+\Lambda ^{D}-A_{s}\right] ^{2}+4A_{s}\Lambda ^{R}} 
}{2A_{s}} , \\
&&\text{where} \\
A_{C} &=&\left( 1-p_{C}^{R}\right) B+D\text{ \ and }A_{N}=\left(
1-p_{N}^{R}\right) B+C+D.
\end{eqnarray*}

Figure 4 shows an example of mortality surfaces for cooperators and
noncooperators for a huge value of $D$. Figure 5 shows sections of the
mortality surfaces for the specific values of the assortments probabilities $
p_{s}^{R}$, where $p_{C}^{R}<p_{N}^{R}$, and the resulting stable role
distributions $\tilde{q}_{s}^{D}$. It shows that for the same switching
rates $\Lambda ^{R}$ and $\Lambda ^{D}$ for both strategies cooperators have
greater mortality.\bigskip

However, the situation changes when we relax the assumption of neutral
switching rates. Figure 6 shows this situation when we have different
switching rates $\Lambda _{s}^{R}$ and $\Lambda _{s}^{D}$ for competing
strategies. The only difference is $\Lambda _{C}^{R}>\Lambda _{N}^{R}$,
which means that cooperators, exhausted by stressful heroic acts, are more
likely to move inside the area and take a rest. Then, the cooperation may
spread even in the case under negative assortment (when $p_{C}^{R}<p_{N}^{R}$
). A similar situation is completely impossible under classical Hamilton's
rule based on couterfactual method.\bigskip 

\begin{figure}[!h]
\centering
\includegraphics[width=10cm]{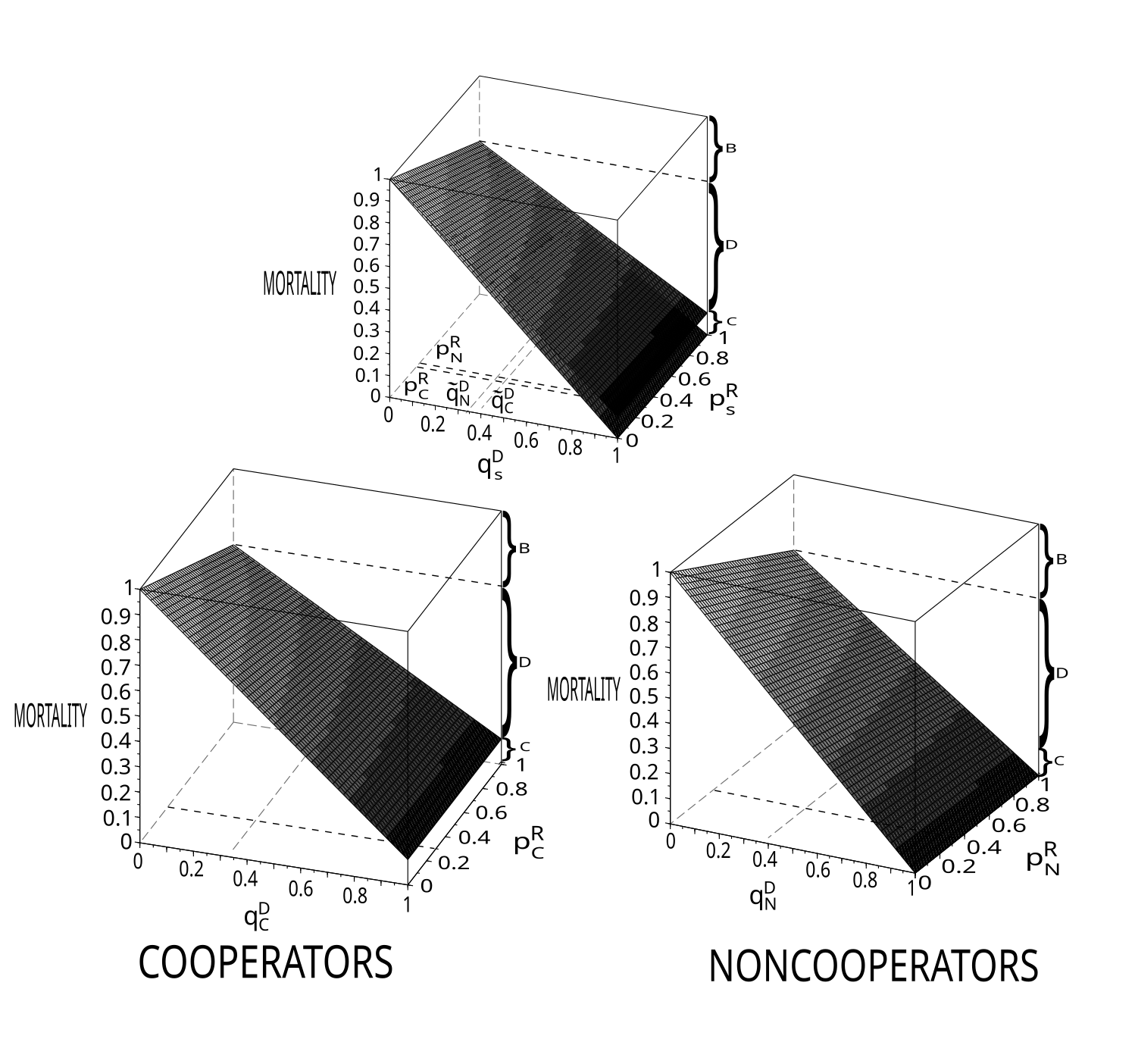} 
\caption{\emph{The examples of\ the mortality surfaces for cooperators and
noncooperators for parameters }$C=0.1$, $B=0.3$, $D=0.6$.}
\end{figure}

\begin{figure}[!h]
\centering
\includegraphics[width=10cm]{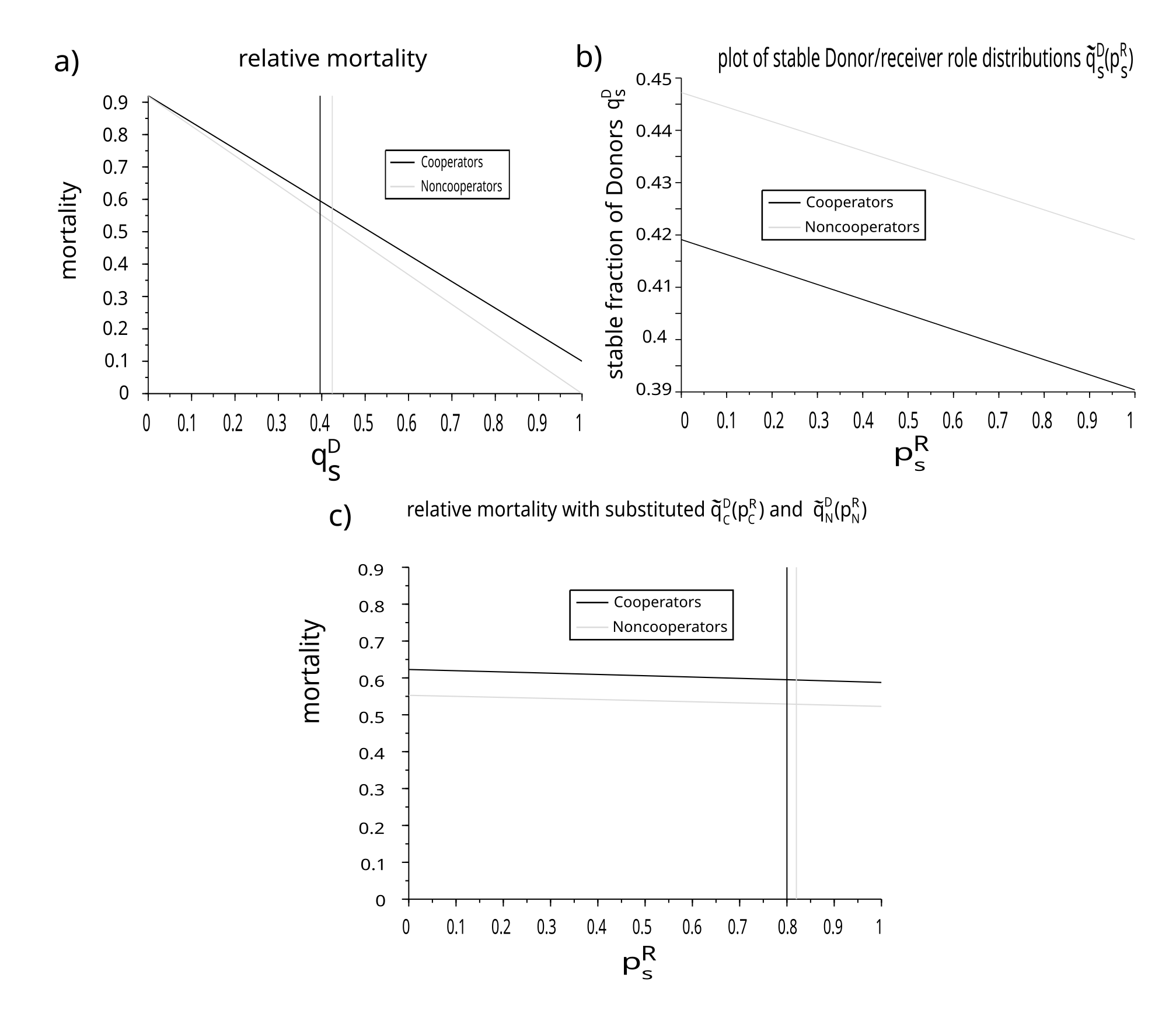} 
\caption{\emph{Plots of mortality and stable role distributions for
parameters $B=0.1$, $C=0.1$, $p_{C}^{R}=0.8$, $p_{N}^{R}=0.82$, $\Lambda
^{R}=0.2$, $\Lambda ^{D}=0.8$.This is the case with negative assortment when noncooperators win. Resulting stable role distributions are $\tilde{q}_{C}^{D}=0.5951506$ and $\tilde{q}_{N}^{D}=0.5285572$.   Panel a) shows
sections along the stable role distribution. Panel b) shows the plot of the
stable role distributions for both strategies. Panel c) shows sections along
the assortment probabilities with substituted functions from panel b).}}
\end{figure}

\clearpage

\begin{figure}[!th]
\centering
\includegraphics[width=10cm]{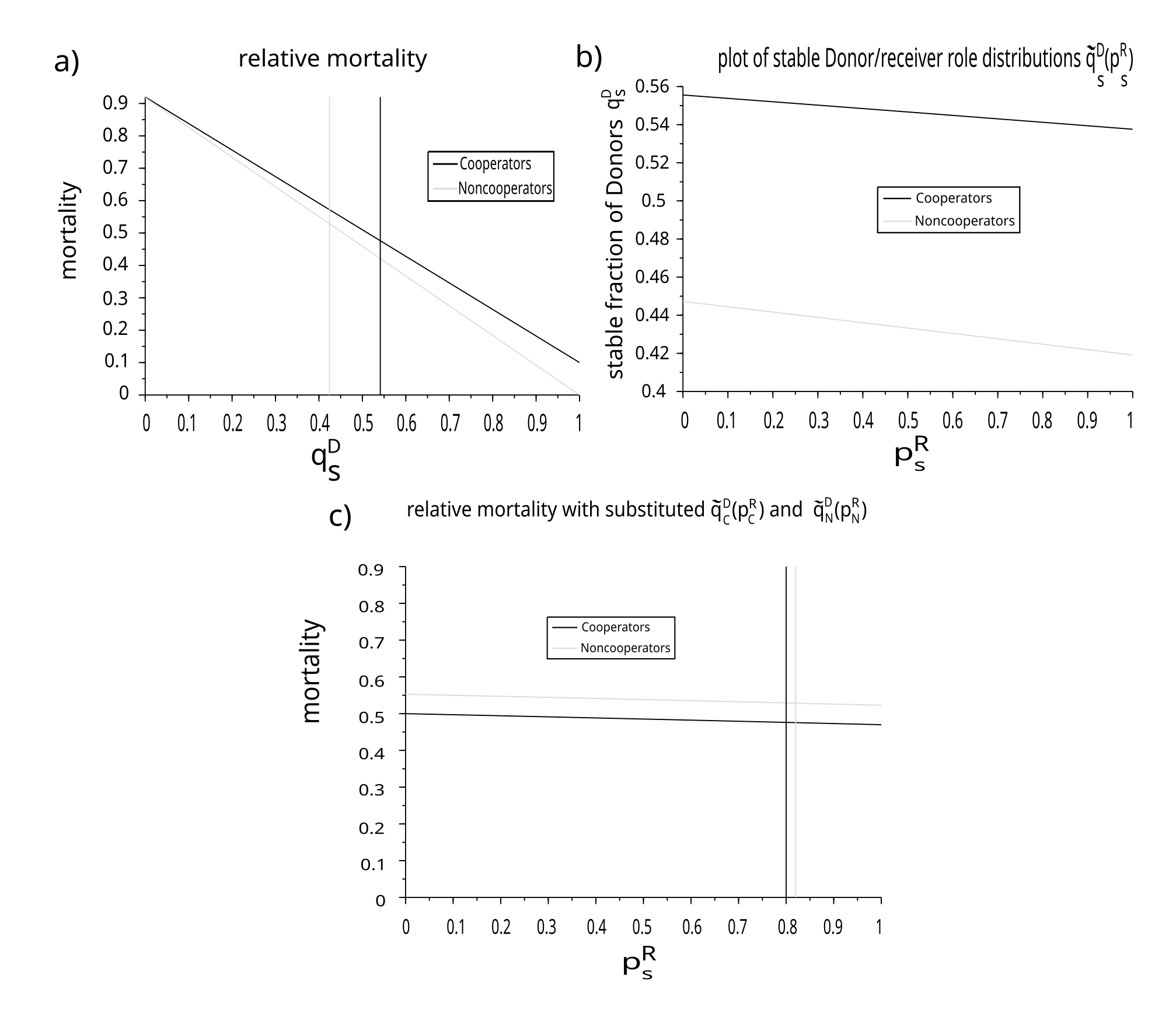}  
\caption{\emph{\ Plots of mortality and stable role distributions for
different strategy specific role switching rates $\Lambda _{C}^{R}=0.5$, $
\Lambda _{C}^{D}=0.8$, $\Lambda _{N}^{R}=0.2$ and $\Lambda _{N}^{D}=0.8$. Other
parameters are the same as in Fig. 5. Resulting stable role distributions
are $\tilde{q}_{C}^{D}=0.4761872$ and $\tilde{q}_{N}^{D}=0.5285572$. \ In
this case cooperative strategy has smaller mortality despite $
p_{C}^{R}<p_{N}^{R}$.}}
\end{figure}

\bigskip

\subsection{Case of $D<0$}

\begin{figure}[!h]
\centering
\includegraphics[width=11cm]{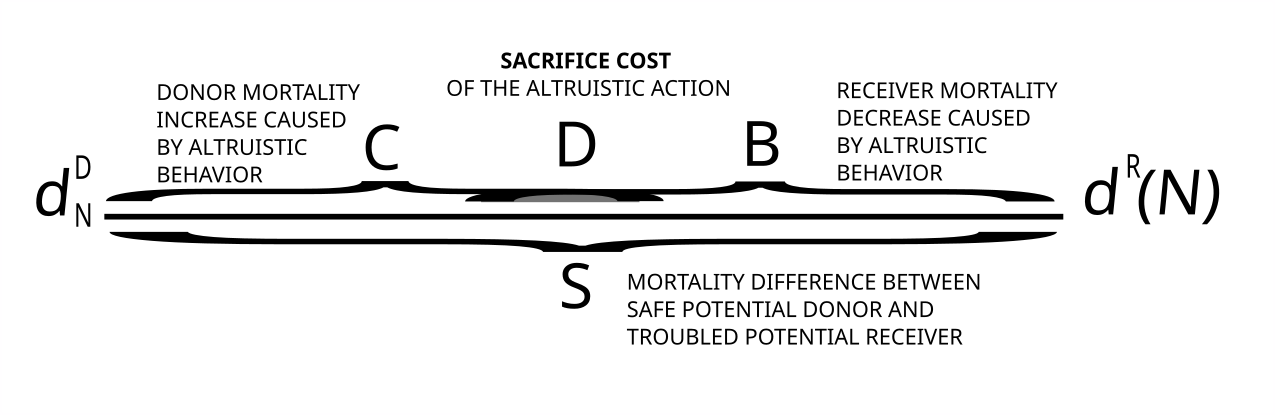}
\caption{\emph{\ Case of the overlapping fitness effects, when donor suffers
greater mortality than helped receivers.}}
\end{figure}
\emph{\ }

A negative value of $D$ may occur when $C+B>S$, which means that changes in
the values of the mortalities caused by altruistic action overlap and in
effect inverts the inequality between the values of the mortalities of donor
and receiver (this is depicted in Figure 7). Then, parameter $D$ can be
termed the \textbf{cooperator's sacrifice cost} since it acts negatively.

\bigskip

\begin{figure}[]
\centering
\includegraphics[width=10cm]{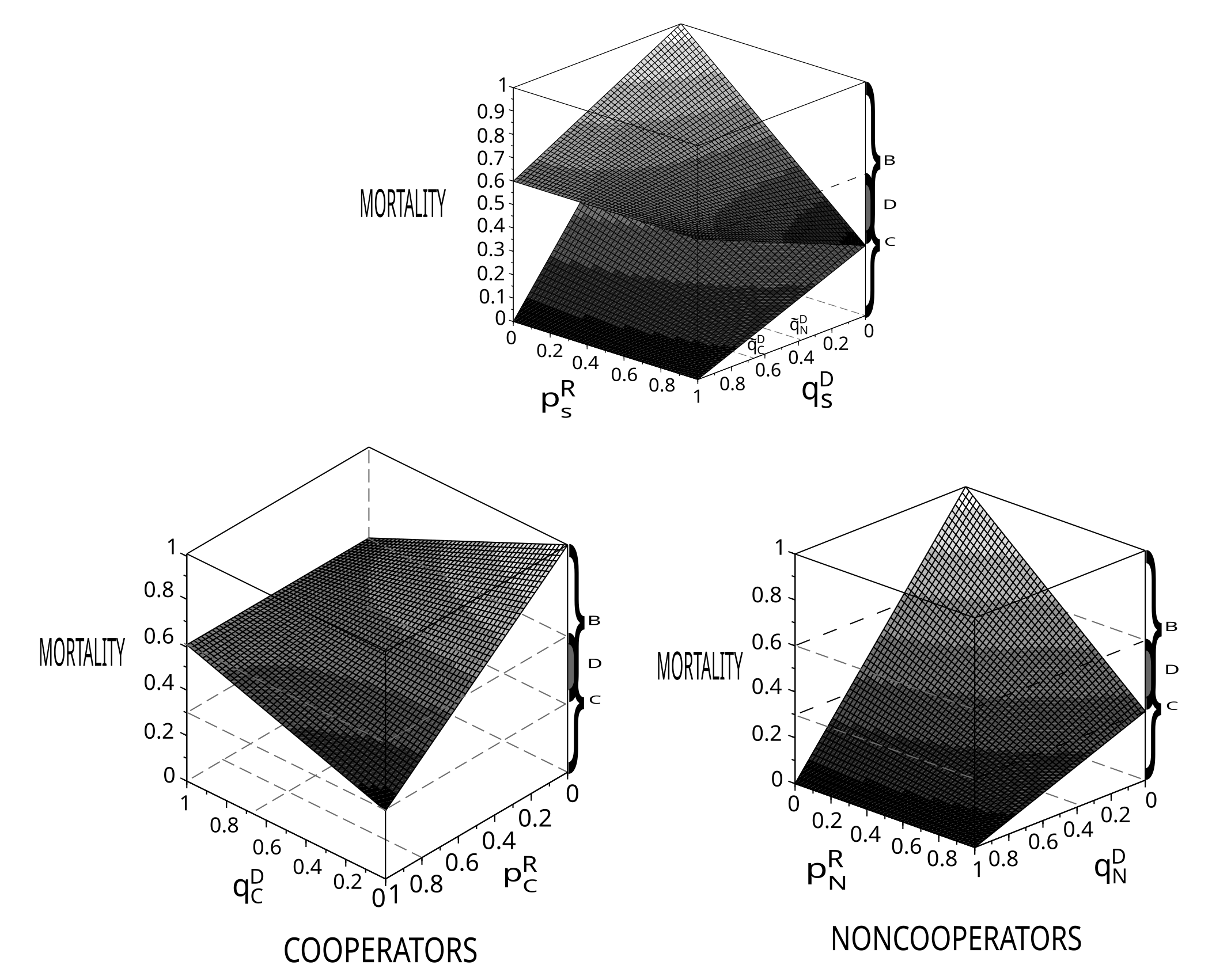}
\caption{ \emph{mortality surfaces for cooperators and noncooperators for
parameters }$C=0.6$, $B=0.7$, $D=0.3$, $p_{C}^{R}=0.8$, $\ p_{N}^{R}=0.1$ }
\end{figure}

\bigskip

This situation may occur when the cooperator cannot hide faster than
receivers and the warning signal exposes him to predator more than assorted
receivers. Let us incorporate the negativity of $D$ into the rule (\ref
{coop2}). Then the general rule for cooperation has the form 
\begin{equation*}
\left[ \left( 1-q_{N}^{D}\right) \left( 1-p_{N}^{R}\right) -\left(
1-q_{C}^{D}\right) \left( 1-p_{C}^{R}\right) \right] B-\left[
q_{C}^{D}-q_{N}^{D}\right] D>q_{N}^{D}C
\end{equation*}

\begin{figure}[!h]
\centering
\includegraphics[width=10cm]{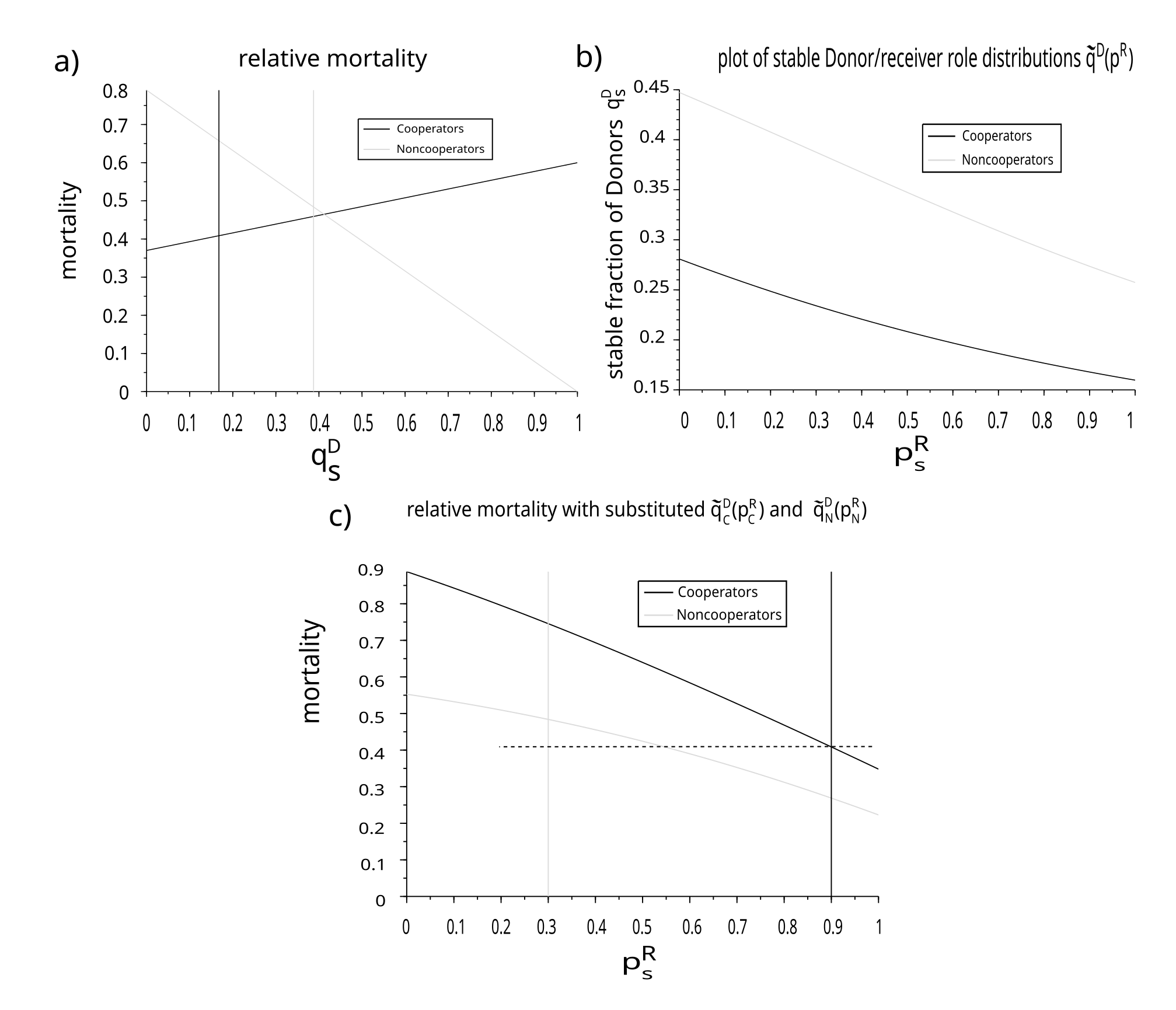} 
\caption{ \emph{Plots of mortality and stable role distributions for
parameters $B=0.7$, $C=0.6$, $p_{C}^{R}=0.9$, $p_{N}^{R}=0.3$ and
strategically neutral role switching rates $\Lambda ^{R}=0.2$, $\Lambda
^{D}=0.8$. Resulting stable role distributions are $\tilde{q}_{C}^{D}=0.4086104$ and $\tilde{q}_{N}^{D}=0.4838735$. Cooperators  win, however the difference in mortality is relatively small despite big difference in assortment probabilities. } }
\end{figure}

Figure 8 shows the example plots of mortality surfaces. In the case with the
same switching rates for both strategies, parameter $D$ acts negatively (Fig.
9) and despite strong assortment probability difference resulting mortality difference is relatively small. When we allow for strategy-specific role switching rates, the situation also changes.
 In this case, when we assume different swithching rates cooperative strategy may also spread despite $p_{C}^{R}<p_{N}^{R}$. However, in this case we need greater flux of cooperators into donor role (Fig. 10). \bigskip

\begin{figure}[!h]
\centering
\includegraphics[width=10cm]{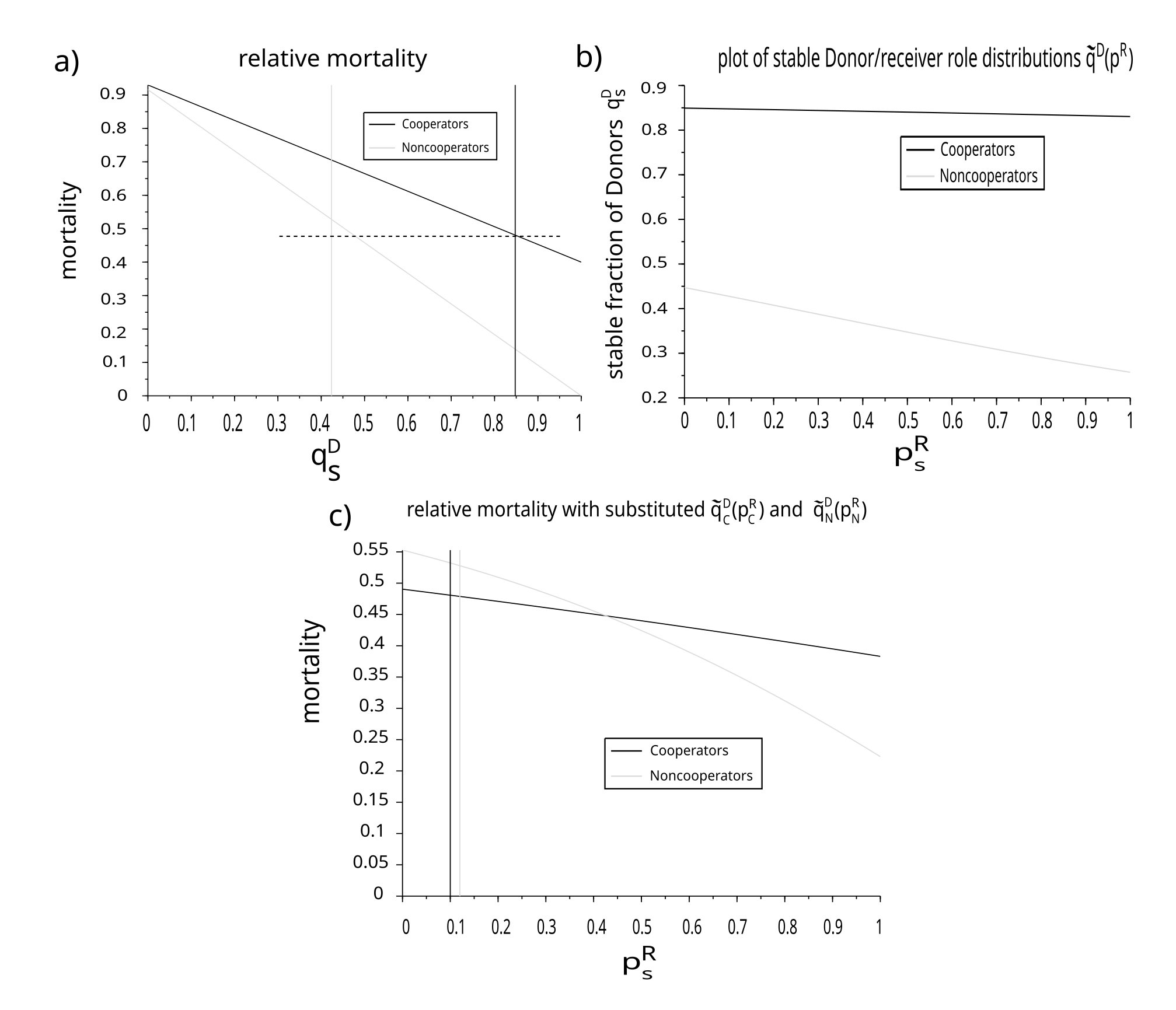} 
\caption{ \emph{\ Plots of mortality and stable role distributions for
parameters $B=0.7$, $C=0.4$, assortment probabilities $p_{C}^{R}=0.1$, $
p_{N}^{R}=0.12$ and different role switching rates for both strategies $
\Lambda _{C}^{R}=4$, $\Lambda _{C}^{D}=0.8$, $\Lambda _{N}^{R}=0.2$, $
\Lambda _{N}^{D}=0.8$. Resulting stable role distributions are $\tilde{q}
_{C}^{D}= 0.4807738$ and $\tilde{q}_{N}^{D}=0.5279256$. Cooperative strategy wins despite negative assortment. } }
\end{figure}

%\clearpage

\subsection{Case of $D=0$}

\begin{figure}[!h]
\centering
\includegraphics[width=11cm]{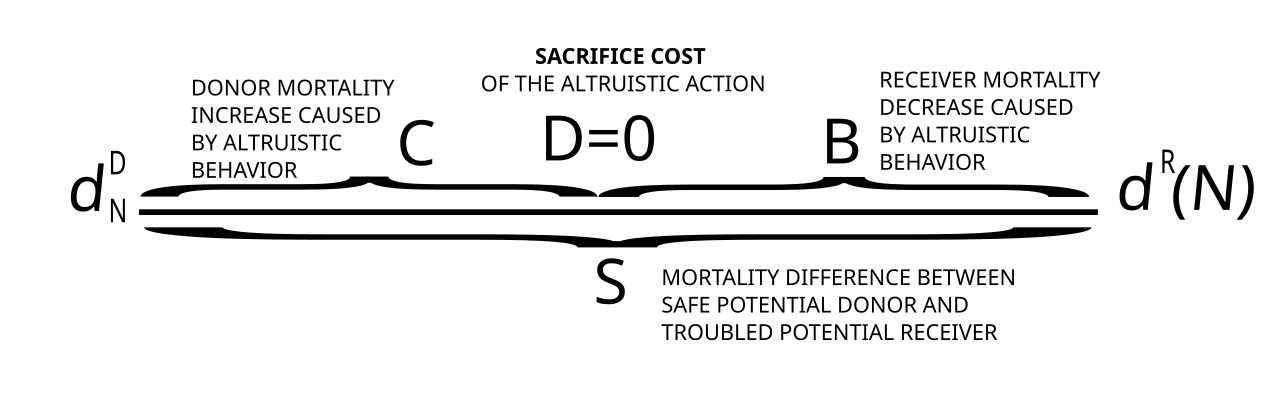}
\caption{\emph{\ Fitness effects resulting from the mortality differences
when} $D=0$ }
\end{figure}

Note that for $D=0$, the rule 
%Editor: Please ensure that the intended meaning has been maintained in this edit.
has the form (see Figure 11 for intuitive interpretation) 
\begin{equation}
\frac{\left[ \left( 1-q_{N}^{D}\right) \left( 1-p_{N}^{R}\right) -\left(
1-q_{C}^{D}\right) \left( 1-p_{C}^{R}\right) \right] }{q_{N}^{D}}B>C.
\label{NewHam}
\end{equation}

This case lies between cases from previous subsections. This situation may
happen when the warning signal does not attract the attention of the
predator directly to the cooperative donor, but hidden assorted individuals
can be detected with the same probability. Therefore, the risk is the same
for donor and receivers. Figures 11 and 12 show fitness effects and
mortality surfaces for $B=0.7$ and $C=0.3$. The case of $D=0$ lies between previous cases and produce similar behavior, thus section figures are redundant. 

\bigskip

Summarizing, we obtain a formula expressed in terms of the benefit, the
cost, and the cooperator's survival surplus/sacrifice cost. The last
parameter can describe important biological factors. The survival
surplus/sacrifice cost can arise in many types of problems, for example, in
engaging in the fight to save another individual. A passive individual is
safer than all individuals involved in the fight. For example, in the
problem of the predator warning signal, this parameter 
%Editor: Please ensure that the intended meaning has been maintained in this edit.
may be zero because a selfish individual who spots a predator hides; thus,
it behaves like individuals warned by a cooperator. However, we can imagine
cases in which hidden noncooperators may have higher survival when all other
individuals are exposed and attract the attention of the predator than when
everybody is hidden and has the same risk of being caught.

\begin{figure}[]
\centering
\includegraphics[width=10cm]{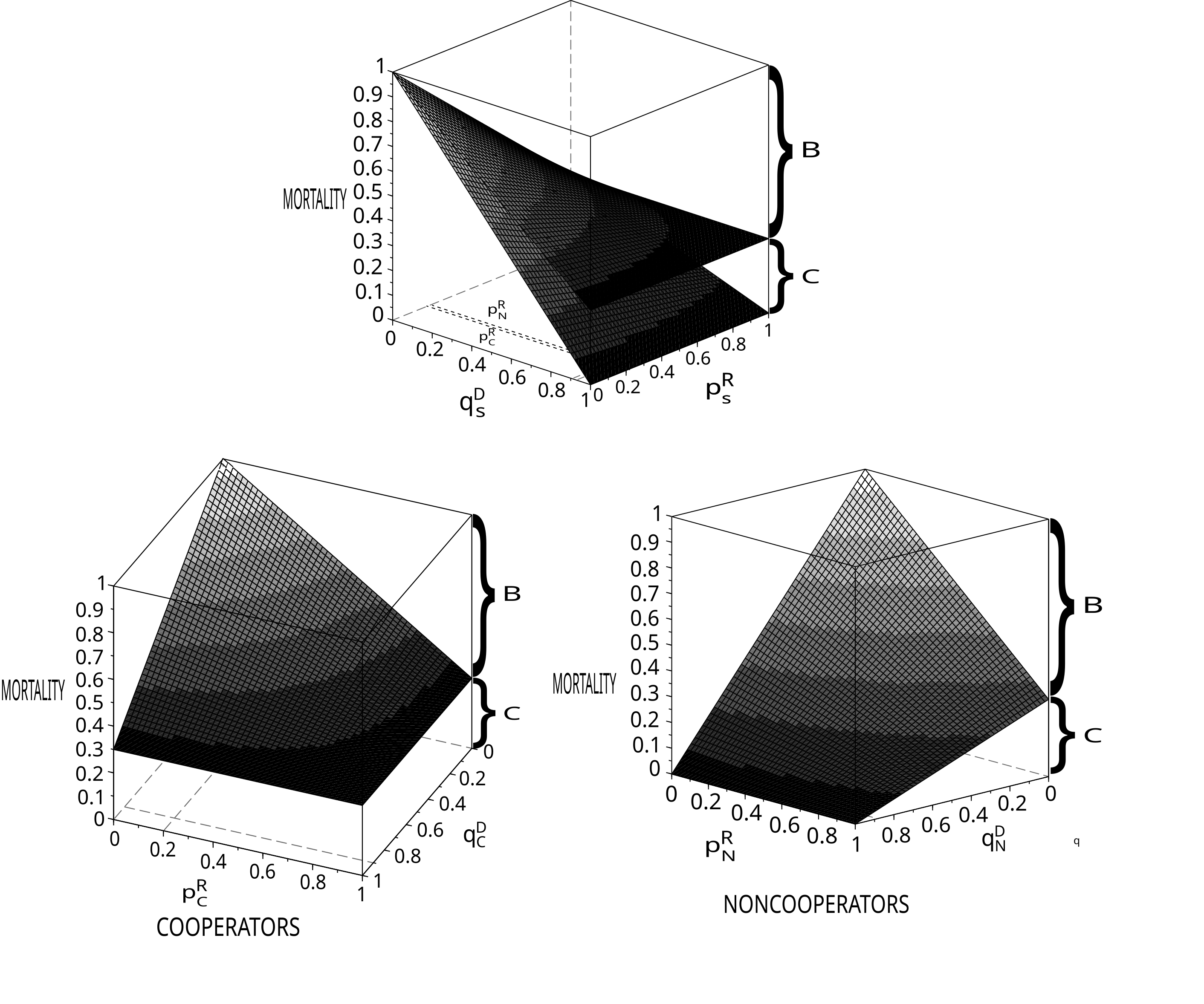}
\caption{ \emph{mortality surfaces for cooperators and noncooperators. Panel
a) for parameters } $C=0.3$, $B=0.7$. }
\end{figure}

\bigskip

The formula (\ref{NewHam}) takes into account the asymmetry in the
distribution of roles. The values of $q_{C}^{R}$ and $q_{N}^{R}$ can be
calculated from the equilibria of the equations (\ref{q-C},\ref{q-N}). By
substituting the obtained equilibria of the switching dynamics we can derive
the function describing the fitness approximation in the long term, based on
the assumption that the switching dynamics is sufficiently close to the
equilibrium. We do this for the the simplest case of $D=0$ and the simplest
form of the switching dynamics (\ref{q-C}) and (\ref{q-N}) where switching
rates $\Lambda ^{R}$ and$\ \Lambda ^{D}$ are constants. In effect, we obtain
the following formula (detailed derivation in Appendix 9b): 

\begin{gather}
	\left[ p_{C}^{R}-p_{N}^{R}\right] B>  \notag \\
	\frac{\sqrt{\left[ \Lambda ^{R}+\Lambda ^{D}-\left( 1-p_{N}^{R}\right) B-C
			\right] ^{2}+4\Lambda ^{R}\left( \left( 1-p_{N}^{R}\right) B+C\right) }-
		\left[ \Lambda ^{R}+\Lambda ^{D}-\left( 1-p_{N}^{R}\right) B-C\right] }{
		2\left( \left( 1-p_{N}^{R}\right) B+C\right) }\left[ \left(
	1-p_{N}^{R}\right) B+C\right]   \notag \\
	-\frac{\sqrt{\left[ \Lambda ^{R}+\Lambda ^{D}-\left( 1-p_{C}^{R}\right) B
			\right] ^{2}+4\Lambda ^{R}\left( \left( 1-p_{C}^{R}\right) B\right) }-\left[
		\Lambda ^{R}+\Lambda ^{D}-\left( 1-p_{C}^{R}\right) B\right] }{2\left(
		\left( 1-p_{C}^{R}\right) B\right) }\left( 1-p_{C}^{R}\right) B
\end{gather}

Then, the complexity dramatically increases compared to the classical
Hamilton's rule $\left( p_{C}^{R}-p_{N}^{R}\right) B>C$ even in this
simplest possible case. In more realistic cases where switching rates are
functions describing the mechanism responsible for role switching, the
situation can be even more complex. However, note that the underlying
dynamic model is still tractable and can be used instead of a static
approach leading to the complex rules.

\subsection{The kin selection case}

Assume that the group of assorted individuals consists of kins with
relatedness $r$. Recall from the previous sections that in the case of kin
selection, $p_{C}^{R}$ and $p_{N}^{R}$ is replaced by $p_{C}^{kin}$ and $p_{N}^{kin}$ (\ref{pkin}), since cooperators pay the cost only for their
kin. Thus, in the switching dynamics (\ref{q-C}) and (\ref{q-N}), the terms 
$C+p_{C}^{R}B$ and $p_{N}^{R}B$ should be replaced by $C+p_{C}^{kin}B$ for
cooperators and $p_{N}^{kin}B$ for noncooperators. Recall that 
\begin{equation}
p_{C}^{kin}=r(1)+(1-r)g_{C}\text{ \ \ \ \ and \ \ \ \ \ }
p_{N}^{kin}=r(0)+(1-r)g_{C}
\end{equation}
Then the fractions of unhelped individuals will be (derivation in Appendix
10): 
\begin{eqnarray*}
\left( 1-q_{N}^{D}\right) \left( 1-p_{N}^{kin}\right)  &=&\left(
1-q_{N}^{D}\right) \left( 1-(1-r)g_{C}\right)  \\
\left( 1-q_{C}^{D}\right) \left( 1-p_{C}^{kin}\right)  &=&\left(
1-q_{C}^{D}\right) (1-r)\left( 1-g_{C}\right) 
\end{eqnarray*}
and the resulting $B$ bracketed term will be 
\begin{eqnarray*}
&&\left( 1-q_{N}^{D}\right) \left( 1-p_{N}^{kin}\right) -\left(
1-q_{C}^{D}\right) \left( 1-p_{C}^{kin}\right)  \\
&=&\left[ q_{C}^{D}-q_{N}^{D}\right] \left( 1-(1-r)g_{C}\right) +\left(
1-q_{C}^{D}\right) r
\end{eqnarray*}
Therefore, (\ref{coop2}) is

\begin{equation}
\left[ \left[ q_{C}^{D}-q_{N}^{D}\right] \left( 1-(1-r)g_{C}\right) +\left(
1-q_{C}^{D}\right) r\right] B+\left[ q_{C}^{D}-q_{N}^{D}\right] D>q_{N}^{D}C,
\label{kincoop}
\end{equation}
and the form revealing the impact of the parameter $r$ is 
\begin{equation}
\left[ \left[ q_{C}^{D}-q_{N}^{D}\right] \left( 1-g_{C}\right) +r\left( 1- 
\left[ \left( 1-g_{C}\right) q_{C}^{D}+g_{C}q_{N}^{D}\right] \right) \right]
B+\left[ q_{C}^{D}-q_{N}^{D}\right] D>q_{N}^{D}C
\end{equation}
In the case of $D=0$, the formula (\ref{kincoop}) simplifies to 
\begin{equation}
\frac{\left[ q_{C}^{D}-q_{N}^{D}\right] \left( 1-(1-r)g_{C}\right) +\left(
1-q_{C}^{D}\right) r}{q_{N}^{D}}B>C.  \label{kinhamD0}
\end{equation}
$\newline
$Therefore the success depends on the frequency of cooperators in the
population described by $g_{C}$. Only in the case when $
q_{N}^{D}=q_{C}^{D}=q^{D}$ does the above formula reduce to the Hamilton's
rule (\ref{HamKin}) 
\begin{equation*}
r\frac{\left( 1-q^{D}\right) }{q^{D}}B>C.
\end{equation*}

For which values of $g_{C}$ rule (\ref{kinhamD0}) will be satisfied?\bigskip

THEOREM 1

Formula (\ref{kinhamD0}) is satisfied when:

a) For $\left[ q_{C}^{D}-q_{N}^{D}\right] >0$ 
\begin{eqnarray*}
g_{C} &<&\tilde{g}_{C}=\frac{q_{C}^{D}+\left( 1-q_{C}^{D}\right)
r-q_{N}^{D}\left( 1+\frac{C}{B}\right) }{\left( q_{C}^{D}-q_{N}^{D}\right)
(1-r)} \\
\tilde{g}_{C} &>&0\text{ \ \ \ \ when \ \ \ \ }q_{N}^{D}<\frac{
q_{C}^{D}+\left( 1-q_{C}^{D}\right) r}{\left( 1+\frac{C}{B}\right) } \\
\tilde{g}_{C} &<&1\text{ \ \ \ \ when \ \ \ \ }q_{N}^{D}>\frac{r}{r+\frac{C}{
B}} .
\end{eqnarray*}

Then $0<\tilde{g}_{C}<1$ if 

\begin{eqnarray*}
	\frac{r}{r+\frac{C}{B}} &<&q_{N}^{D}<\frac{q_{C}^{D}+\left(
		1-q_{C}^{D}\right) r}{\left( 1+\frac{C}{B}\right) } \\
	q_{C}^{D} &>&\frac{r}{r+\frac{C}{B}} .
\end{eqnarray*}

Thus cooperation spreads when $\tilde{g}_{C}\geq 1$, which happens when $
q_{N}^{D}\leq \frac{r}{r+\frac{C}{B}}$.

b) For $\left[ q_{C}^{D}-q_{N}^{D}\right] <0$ 
\begin{eqnarray*}
g_{C} &>&\tilde{g}_{C}=\frac{q_{C}^{D}+\left( 1-q_{C}^{D}\right)
r-q_{N}^{D}\left( 1+\frac{C}{B}\right) }{\left( q_{C}^{D}-q_{N}^{D}\right)
(1-r)} \\
\tilde{g}_{C} &>&0\ \ \ \text{when}\ \ \ \ \ \ q_{N}^{D}>\frac{
q_{C}^{D}+\left( 1-q_{C}^{D}\right) r}{\left( 1+\frac{C}{B}\right) } \\
\tilde{g}_{C} &<&1\text{ \ \ \ \ when \ \ \ }q_{N}^{D}<\frac{r}{r+\frac{C}{B}
} .
\end{eqnarray*}
Then $0<\tilde{g}_{C}<1$ if 

\begin{eqnarray*}
	\frac{r}{r+\frac{C}{B}} &>&q_{N}^{D}>\frac{q_{C}^{D}+\left(
		1-q_{C}^{D}\right) r}{\left( 1+\frac{C}{B}\right) } \\
	q_{C}^{D} &<&\frac{r}{r+\frac{C}{B}} .
\end{eqnarray*}

Thus cooperation spreads when $\tilde{g}_{C}\geq 1$, which happens when $
q_{N}^{D}\geq \frac{r}{r+\frac{C}{B}}$.

Proof in Appendix 11\bigskip

Therefore, there are possible situations that cooperators can dominate the
population, or noncooperators can win. However, we have a third scenario.
Then if $\left[ q_{C}^{D}-q_{N}^{D}\right] >0$, then polymorphic state $\tilde{g}_{C}$ can be stable if
exists, but cooperative strategy cannot dominate the population. On the \
other hand, if $\left[ q_{C}^{D}-q_{N}^{D}\right] <0$ \ then the monomorphic
cooperative population is stable, but the rare cooperative strategy cannot
successfully invade the population. Therefore, when $\left[
q_{C}^{D}-q_{N}^{D}\right] >0$ , cooperators can spread in the population
until reaching a stable mixed state. Strategies can compete by adjusting
their mobility patterns determining switching rates. If parameters change,
leading to $\left[ q_{C}^{D}-q_{N}^{D}\right] <0$, then the state $\tilde{g}
_{C}$ becomes unstable. In effect, cooperators can dominate the population
or die out due to random fluctuation of the population state.

\section{Discussion}

We combined the demographic game approach (with explicit mortality and
fertility payoffs) with Hamilton's rule based on couterfactual method. The
model parameterized by survival probability differences of the critical
event was derived. In effect, the model can be parameterized by empirically
observable parameters and does not lead to the calculus of the unborn
offspring. Therefore, it can act as a predictive model. The framework was
applied to the predator alarm call problem. An important aspect of this type
of problems, which cannot be modeled by standard game theory, is that we
have two roles. The first role is the active donor exhibiting the strategy,
while the second is the passive receiver whose strategy is latent. The
assumption that receivers are passive and their strategies latent is crucial
from a biological point of view. "Additivity" of payoffs (also called "equal
gains from switching" property) is the logical result of this assumption.
The cost for the donors is not the result of the receivers' strategy but
from the external threat that affects the receiver and thus the helping
donor. 
%Editor: Please ensure that the intended meaning has been maintained in this edit.
The "nonadditive" cases assume that, for example, donors provide different
benefits for different strategies, which contradicts the receiver passivity
assumption. This implies the need for external mechanisms to discern
strategies and identify non-cooperative individuals. The classical
formulation works well for cases when roles are independently drawn at every
focal interaction event, and not only for pairwise interactions. A good
example of a problem of this type is Haldane's anecdotal claim
\textquotedblleft \textit{I would gladly give up my life for two brothers or
eight cousins}\textquotedblright , describing the provision of help for a
drowning individual. However, in some problems, this may not be the case.
Then we need the generalization described in the main result of the paper.

\subsection{Main result: Rule for cooperation under state distribution
asymmetry}

These complicated cases can be described by models extended by equations
describing the role-switching dynamics \cite{Argasinski and Rudnicki 2021}. An
important result shown by the new framework is that different strategies may
have different role distributions. The resulting equilibria of the role
distributions (if they exist) should be considered in the general rule for
cooperation, describing the generalization of the classical theory. However,
even in the simplest case of switching dynamics with constant switching
rates, the substitution of the calculated equilibria to the selection rule
lead to extremely complicated nonlinear condition. 
%Editor: Please ensure that the intended meaning has been maintained in this edit.

We thus obtain the general condition affected by differences in role
distributions, which may be termed \textit{state distribution asymmetry}. In
addition to the classical components describing the cost and benefit, the
resulting condition contains a third component. This component may have
different interpretations depending on its value. If cooperative donors'
mortality is smaller than the mortality of the helped receivers, and it can
be termed the \emph{survival surplus}, and it should be added to the actual
benefit in the general rule for cooperation. In the second case, when
cooperative donors' mortality is greater than the mortality of the helped
receivers, therefore, the component can be termed the \emph{sacrifice cost}.
This component should be subtracted from the actual benefit in the general
rule for cooperation. When role switching rates are the same for both
competing strategies, the impact of the third parameter $D$ is negative.
However, when we allow for different switching rates for different
strategies we can obtain cases when cooperation may spread in the cases when
assortment mechanism is not efficient (i.e. probability of receiving help is
slightly greater for noncooperators than cooperators). This is completely
impossible under classical Hamilton's rule. Especially interesting is the
case of \emph{\ sacrifice cost}. Then, for small values of assortment
probabilities, cooperation can spread when the assortment mechanism is not
effective. However, for values of the cooperators assortment probability closer to 1,
the noncooperators assortment probability should be significantly smaller to allow the spread of the
cooperative strategy. Therefore, we have cases when strong  \emph{
Machiavellian intelligence }\cite{Gavrilets and Vose 2006, Waal 2007},
responsible here for the recognition of free-riding non-cooperators, is not
necessary. However, there are other cases when it should be efficient. In
addition, in the new model, the kin selection case is more complicated. The
limitation of the altruistic action to kins does not protect against the
impact of the current cooperative gene frequency in the population, as in
classical Hamilton's rule.  \bigskip

\subsection{Future extensions of the proposed methodology}

New model presented in this paper still contains certain simplifications
inherited from the existing state of the art. One simplification is lack of
the explicit description of the assortment mechanism. The assortment
probabilities are simple constants. We can imagine that the dynamics of
underlying mechanism based on strategy recognition may depend on the
strategy frequencies. This may be especially interesting for the sacrifice
cost\ case, when we observed different outcomes for low and high assortment
probabilities. This is related to the problem of elimination of free-riding
noncooperators. Correctly recognized free riders are not helped; however, in
some cases, such as predator alarm signals, they can benefit from the
general cooperative action toward other cooperators. In these cases,
recognized free riders should be expelled from the group (or even killed).
This aspect is another open question resulting from the new framework.

In general, the distribution of roles resulting from the selection
mechanisms may be an important tool for explaining many biological
phenomena. For example, help for a sick individual may take the form of
supporting her with necessary supplies but may not cure her. If this
individual suffers from an infectious disease, then the altruistic action
may lead to infection of the cooperative donor, and as a result, the
fraction of the strategy carriers finding themselves in trouble may
increase. Thus we have another problem with different switching rates for
different strategies. This is interesting from the point of view of the
latest evidence showing that infected vampire bats avoid other members of
their group \cite{Ripperger et al 2020}. In addition, role switching dynamics may
play an important role in the evolution of the social structure and the
division of labor among social insects \cite{Wilson and Holldobler 2005, Thompson 2006, Thompson et al 2013}. Empirical evidence for \emph{C.floridanus} ants show the sophisticated role-switching mechanism between
tenders and foragers \cite{Tripet and Nonacs 2004}. Switching to a risky forager
role is related to the age of the individual (older ants forage more
likely), while a switch to a less risky role is random. Thus, in this case,
switching dynamics are affected by age structure. In addition, ants are
focused on a particular activity, such as foraging, for a longer time, and
switching to other activities is conditional on the behavior of other
individuals.\ This can be modeled by a combination of age-structured models
\cite{Argasinski and Broom 2021} with additional role-switching dynamics
\cite{Argasinski and Rudnicki 2021}. This is another potential direction for
future research. We should also be aware that the new framework (as well as
the basic evolutionary game theoretic framework) ignores the population
genetics aspects, which can be important in problems related to kin
selection \cite{Garay et al 2018a, Garay et al 2018b, Garay et al 2019, Garay et al 2024, Su et al 2022, Allen et al 2024}. However, the same problem occurs in general evolutionary game theory,
which in most cases assumes asexual clonal reproduction. Some genetic
details can be introduced via payoff functions, such as in multi-population
sex ratio models \cite{Argasinski 2012, Argasinski 2013, Argasinski 2018}. However, the development of
the general synthesis between evolutionary games and population genetics is
still not sufficiently addressed. 

\section*{Appendix 1 Hamilton's Rule (counterfactual method)}

In the frequency-dependent evolutionary game structure, survivors of the
focal interaction split up, and the lonely look for another random
encounter. If the frequency of cooperators is low, then the chance of
receiving help from another cooperator is small. However, when the
assumption of a panmictic population is relaxed, cooperative individuals can
perform altruistic actions (with an associated cost) toward individuals that
can be recognized by some observable cues.\ Some authors assumed the
existence of some assortment mechanism making pairing of cooperators with
cooperators more likely \cite{McElreath and Boyd 2008, Fletcher and Doebeli 2008}. We do not specify here the underlying mechanism (kinship,
reciprocity, etc.) determining the subjects of altruism. However, the
recognition rule is uncertain since individuals do not always exhibit the
cues revealing their strategy. Individuals may be familiar with the
cooperative individual according to the assortment mechanism operating at
the population level. They can simply follow the confirmed cooperator and
support him when he acts as the receiver or abandon the non-cooperator. This
leads to the aggregation of cooperative groups, where the probability of
being helped is significantly greater than that resulting from purely random
encounters.\ The importance of clustering for the spread of cooperation was
also demonstrated by models of structured populations and games on graphs
\cite{Perc et al 2013, Ross et al 2015, Thompson et al 2013, Broom et al 2020}.
Therefore we can define the assortment probabilities $p_{C}^{R}$ and $
p_{N}^{R}$\ of being covered by some neighboring cooperating donor, which
may result from some population mechanisms. Then the condition for the
greater growth of cooperators than non-cooperators is described by a general
formula:    
\begin{equation}
\left[ p_{C}^{R}-p_{N}^{R}\right] B>C,  \label{HamiltonsRule}
\end{equation}
where cost $C$\ and benefit $B$\ describe the fitness effects of the
altruistic action on donor and receiver and $p_{C}^{R}$ and $p_{N}^{R}$\ are
probabilities of receiving help for both strategies. This formula (or
similar ones) can be found in many papers (for example, \cite{McElreath and Boyd 2008, Fletcher and Doebeli 2008, Alger and Weibull 2012, Okasha and Martens 2016a}). It means that the cost interpreted as the donor's mortality increase
should be smaller than the sum of the fitness effects resulting from
mortality decreases of the receivers. The above condition can be satisfied
only for $p_{C}^{R}>p_{N}^{R}$, which means that cooperators are more likely
to receive help than defectors, which can be caused by different mechanisms.
Indeed, for the low frequencies of the cooperators, this assortment can be
realized by a very simple mechanism. In this special case, termed \textbf{\
kin selection} interactions are limited to kin only. Thus, instead of
guessing the strategy of the assorted receiver, donors support only kin of
some specific degree (for example, only brothers and sisters or only
cousins). In the general case, we can describe the degree of kinship between
two individuals as the number \ of ancestor generations till the last common
ancestor (then, $r$ is the probability that both actors share the altruist
gene from a common ancestor, hereafter referred to as kin relatedness). The
cooperative donor after kin recognition pays the conditional cost $C$ and
delivers the conditional benefit $B$. However, for different strategies, we
have different conditional probabilities that this potential kin donor is a
carrier of the altruist gene ($p_{C}^{kin}$ and $p_{N}^{kin}$,
respectively). The derivation of these probabilities can be found in
McElreath and Boyd (2008). The difference from (\ref{HamiltonsRule}) is
that in the kin selection case, $p_{s}^{kin}$\ describes the probability of
inheriting the cooperative strategy from some random ancestor, not being
helped by some assorted cooperator as the parameters $p_{s}^{R}$ in formula (\ref{HamiltonsRule}). The receiver carries the same gene from a common
ancestor with probability $r$, but he can also carry this gene from another
source with probability\ proportional to the cooperative gene frequency
(described by the parameter $g_{C}$). Similarly, a kin individual of a
noncooperative receiver does not carry the cooperative gene with probability 
$r$ but can carry it from other sources with probability $g_{C}$. When
we limit interactions to kin with relatedness $r$, then 
\begin{equation}
p_{C}^{kin}=r(1)+(1-r)g_{C}\text{ \ \ \ \ \ \ \ \ and \ \ \ \ \ }
p_{N}^{kin}=r(0)+(1-r)g_{C}  \label{pkin}
\end{equation}
and the brackets $(0)$ and $(1)$ describe multiplication by probabilities $0$
and $1$. Since $p_{C}^{kin}-p_{N}^{kin}=r$, formula (\ref{HamiltonsRule})
becomes 
\begin{equation}
rB>C,  \label{HamiltonsRule2}
\end{equation}
\newline
which is the classical Hamilton's rule. Therefore, the limitation of
altruistic actions to kins is the strategy to overcome the pressure of
frequency-dependent selection. It produces a selective advantage
independently of the cooperative gene frequency in the population. The
disadvantage is that the range of possible cooperation is dramatically
reduced. From the point of view of our panmictic population, it should be
regarded as the evolution of nepotism rather than altruism since it involves
the refusal to help nonkin. This is supported by empirical observations
\cite{Dunford 1977, Sherman 1977, Hoogland 1983}. Note that the condition (\ref
{HamiltonsRule2})) is similar to the condition for the positive growth of
cooperators $C<p_{C}^{R}B$ (when we assume $r_{s}=p_{C}^{R}$), and the
difference is that the parameter $r_{s}$ (probability of identity by state)
is replaced by $r$ (probability of identity by descent). This may be
misleading and potentially can cause misunderstanding (see Appendix 2 for a
detailed discussion on this problem).\bigskip 

\section*{Appendix 2: What is the difference between $C<r_{s}B$ and $C<rB$?}

In addition to popular fallacies \cite{Park 2007, West et al 2011} associated
with Hamilton's rule, there is one popular mistake related to the
relationships between Haamilton's rule and kin selection concepts. The
question is: should the relatedness be defined as the probability that the
receiver is the carrier of the cooperative gene, or should it also be
inherited from the common ancestor? This problem was critically discussed by
Gintis (2013) and can be found, for example, in \cite{Bourke 2021}. Moreover,
Encyclopedia Britannica states that 

``\emph{Relatedness is the probability that a gene in the potential altruist
is shared by the potential recipient of the altruistic behavior}''

\noindent without explicit\ reference to genealogy. Thus in this
case we have $r_{s}=p_{C}^{R}$. The source of the problem is
as follows: the condition for positive growth $C<r_{s}B$ reduces to 
\begin{equation*}
C<r_{s}B=p_{C}^{kin}B=\left[ r+(1-r)g_{C}\right] B
\end{equation*}
for the kin selection case. On the other hand, the general condition for
cooperation $C<\left[ p_{C}^{R}-p_{N}^{R}\right] B$ (\ref{HamiltonsRule}) in
the kin selection case reduces to $C<rB$. Then, the relationships between $
r_{s}$ and $r$ can be summarized as 
\begin{align*}
\boldsymbol{C}\boldsymbol{<r_{s}B}& \simeq p_{C}^{kin}B\text{ \ \ }
\Longrightarrow \text{\ \ \ \ }C<\left( r+(1-r)g_{C}\right) B. \\
C& <\left[ p_{C}^{kin}-p_{N}^{kin}\right] B\text{ \ \ }\Longrightarrow \text{
\ \ \ \ }\boldsymbol{C<}r\boldsymbol{B}.
\end{align*}
Thus, what is the difference between $C<r_{s}B$ and $C<rB$? The condition $
C<r_{s}B$, where $r_{s}$ is the probability that the receiver is the carrier
of the cooperative gene (\textit{identity by state} in terms of population
genetics), is the condition for a positive impact of the act of altruism on
the growth rate of cooperators. Thus, it is not sufficient for the spread of
altruism. On the other hand, the condition $C<rB$, where $r$ is the
probability that the receiver inherited cooperative genes from the common
ancestor (\textit{identity by descent}), is the condition for greater growth
of cooperators over noncooperators. This is the correct condition for
altruism, albeit limited to kin only. This aspect is important from an
educational point of view. Hamilton's rule $C<rB$\ is not obvious and
intuitive without the explanation that it results from condition (\ref
{HamiltonsRule}). When presented alone, the rule 
%Editor: Please ensure that the intended meaning has been maintained in this edit.
can easily \textquotedblleft mutate\textquotedblright\ into condition $
C<r_{s}B$ , which is intuitive but not sufficient for the spread of the
cooperative trait. Thus, the risk of possible misunderstanding is very
high.\ Therefore, textbooks and popular science papers should clearly
explain the whole chain of reasoning of how we derive Hamilton's rule from
the more general condition (\ref{HamiltonsRule}).\bigskip

\section*{Appendix 3 Basic assumptions of event-based modeling and
demographic game approach}

The event-based approach focused on the explicit dynamics of interaction
events in time, and the aggregation of their outcomes was introduced in
\cite{Argasinski and Broom 2013} later extended and clarified in \cite{Argasinski and Broom 2018a} and completed with the derivation of eco-evolutionary stability
conditions in \cite{Argasinski and Broom 2018b}. For the derivation of the growth
equation, we can use the method from \cite{Argasinski and Broom 2018a}. Assume
that individuals are involved in different types of interaction\ events
described by demographic outcomes (mortality and fertility). We can derive
the vital rates (birth and death rates) as the product of interaction rates
and demographic parameters describing the number of offspring and the
probability of death in a single interaction. The general growth equation of
the subpopulation of individuals with strategy $s$ (described by subscript $
s $, while superscript $j$ describes the event type) is 
\begin{equation}
\dot{n}_{s}=n_{s}\sum_{j}\tau ^{j}\left( W_{s}^{j}-d_{s}^{j}\right) ,
\label{basic}
\end{equation}
where

$\tau ^{j}$ is the interaction rate (event occurrence rate) of the $j$-th
type event,

$W_{s}^{j}$ is the fertility payoff (number of offspring) in the $j$-th type
event, and

$d_{s}^{j}$ is the mortality payoff in the $j$-th type event.\newline
The analyzed trait under selection, described by different strategies, may
affect few or even only one type of interaction (we limit our attention to
this case). This interaction is described as the \textit{focal game}
(described by $\tau ^{f}$, $W_{s}^{f}$ and $d_{s}^{f}$). Other types of
events constitute the background fitness, which is the same for all
strategies 
\begin{equation}
R_{b}=\dfrac{\sum_{j\neq f}\tau ^{j}\left( W_{s}^{j}-d_{s}^{j}\right) }{\tau
^{f}}.
\end{equation}
\newline
Some of the background events may depend on the population size; thus, the $
R_{b}$\ parameter may be a function describing the density-dependent effects
(for simplicity, we do not describe this explicitly). This leads to the
basic growth equation 
\begin{equation*}
\dot{n}_{s}=n_{s}\tau ^{f}\left( R_{b}+W_{s}^{f}-d_{s}^{f}\right) ,
\end{equation*}
where $\tau ^{f}$ can be set to 1 by changing the timescale (however, this
is still a substantial "construction element" of the differential equation
producing correct unit of the resulting birth and death rates). For the
subject of our paper, in most basic cases, the altruistic action can be
expressed in terms of the average mortality $d_{s}^{f}$ (or equivalently
survival) of the individual carrying strategy $s$.\ Therefore, the fertility
payoffs will be not present due to the lack of direct fertility outputs
related to those events, leading to 
\begin{equation}
\dot{n}_{s}=n_{s}\left( R_{b}-d_{s}^{f}\right) ,  \label{basic2}
\end{equation}
\ Then equation (\ref{basic2}) can be rescaled to replicator dynamics
coupled with state-switching dynamics \cite{Argasinski and Rudnicki 2021}. In
this paper will use one of the most simple variants of this approach.
Because this is a novel methodology, the state-switching equations will be
carefully re-derived in the Results section. \bigskip 

\section*{Appendix 4: Relationships with matrix game-theoretic payoff
functions and the "additivity" issue}

One of the serious problems raised in the ongoing discussion is the question
of the additivity of payoffs \cite{Levin and Grafen 2019, Levin and Grafen 2021}. Many authors
have claimed that Hamilton's rule models do not work for "nonadditive"
payoff functions \cite{Nowak et al 2010, Van Veelen 2009, Okasha and Martens 2016a}. This has later been clarified \cite{Van Veelen et al 2017} through the
introduction of the distinction between the \textquotedblleft counterfactual
method\textquotedblright\ (originated by Karlin and Matessi \cite{Karlin and Matessi 1983, Matessi and Karlin 1984, Matessi and Karlin 1986} based on the differences in fitness resulting from
different actions (this method is used in our paper) and the more
general but more complicated \textquotedblleft regression
method\textquotedblright\ \cite{ Marshall 2015, Rousset 2015, Okasha 2016} defining relatedness as the regression coefficient. The general claim
\cite{Van Veelen et al 2017} is that for additive models, where the 
\textquotedblleft equal gains from switching \textquotedblright\
property (which means equality of the sums of elements on both diagonals of
the payoff matrix; \cite{Nowak and Sigmund 1990, Wild and Traulsen 2007}) is
satisfied, both methods are equivalent. However, the regression method also
works for nonadditive cases, but the obtained versions of the Hamilton's
rule are not unique. The basic matrix model used in literature is the
donation game with unspecified (positive) cost $c$ and benefit $b$ \cite{Marshall 2015, Van Veelen et al 2017}, which can be presented in the form:

\begin{equation}
\left[ 
\begin{array}{cc}
0 & b \\ 
-c & b-c
\end{array}
\right] .  \label{DonationMatrix}
\end{equation}
\bigskip

In \cite{Van Veelen et al 2017}, it is described as "\emph{the game between two
possible donors}" and is completed by the form exposing "\emph{what the 
\textbf{opponent} gets}": 
\begin{equation}
\left[ 
\begin{array}{cc}
0,0 & b,-c \\ 
-c,b & b-c,b-c
\end{array}
\right] .  \label{Donation2}
\end{equation}
\bigskip

Note that the receiver is not an opponent but a troubled individual who
needs help, and an altruistic act is not a conflict. The above matrices are
additive, which is criticized as a framework limitation. Additivity means
that for the matrix

\begin{equation}
\left[ 
\begin{array}{cc}
a & b \\ 
c & d
\end{array}
\right] ,  \label{matrix}
\end{equation}

we have the property $a+d=c+b$, termed "equal gains from switching". This
implies conditions i) $y=a-b=c-d$ and ii)$\ x=a-c=b-d$. Those conditions
allow for presentation of matrix (\ref{matrix}) in the forms 
\begin{equation}
\text{i) \ }\left[ 
\begin{array}{cc}
a & a-y \\ 
c & c-y
\end{array}
\right] \text{ \ \ \ \ \ \ \ \ \ ii) \ }\left[ 
\begin{array}{cc}
a & b \\ 
a-x & b-x
\end{array}
\right] .
\end{equation}

When we combine conditions i) and ii) and subtract $a$ from all entries,$\ $
we get the matrix 
\begin{equation}
\left[ 
\begin{array}{cc}
0 & -y \\ 
-x & -x-y
\end{array}
\right] ,  \label{matrixEGS}
\end{equation}

and the values $x$ and$\ y$ can be termed row and column effects. Then, row
effect $x$ is the result of the action of the focal agent, and it is
independent of the strategy of the opponent. Similarly, column effect $y$ is
the result of the opponent's action and is independent of the action of the
focal player. This indeed looks artificial from the perspective of the
standard game theory. In Van Veelen (2018) the additivity is described
as:\bigskip

\emph{Equal gains from switching means that the fitness effects (the costs
and benefits) of the social behaviour are independent of who else
contributes, \textbf{and also independent of whether or not the recipient
performs the behaviour}.} \bigskip

However, if the receiver performs the altruistic act, then he becomes the
donor. Terminology and mathematics of non-cooperative games seems to be
inapropriate since it ignores the division between active donors and passive
receivers. Let us analyze the additivity issue from the point of view of our
simple model with an explicit distribution of roles. In the matrix form, it
can be described as the receiver and donor mortality matrices, where the
first (second) row/column describes the noncooperator (cooperator) 
\begin{equation*}
\left[ 
\begin{array}{cc}
d^{R}(N) & d^{R}(N)-B \\ 
d^{R}(N) & d^{R}(N)-B
\end{array}
\right] \text{ \ \ \ and \ \ \ \ \ }\left[ 
\begin{array}{cc}
d_{N}^{D} & d_{N}^{D} \\ 
d_{N}^{D}+C & d_{N}^{D}+C
\end{array}
\right] \text{\ ,}
\end{equation*}
which can be presented in the combined asymmetric form for bimatrix games 
\textbf{\ (entries describe pairs of payoffs for both players, while column
player is the active donor and the row player is the passive receiver)} 
\begin{align*}
\left[ 
\begin{array}{cc}
d^{R}(N),d_{N}^{D} & d^{R}(N)-B,d_{N}^{D}+C \\ 
d^{R}(N),d_{N}^{D} & d^{R}(N)-B,d_{N}^{D}+C
\end{array}
\right] & = \\
(d^{R}(N),d_{N}^{D})\left[ 
\begin{array}{cc}
1,1 & 1,1 \\ 
1,1 & 1,1
\end{array}
\right] & +\left[ 
\begin{array}{cc}
0,0 & -B,C \\ 
0,0 & -B,C
\end{array}
\right] .
\end{align*}
Our matrix takes into account the distinction between roles, and one donor
and one receiver always participates in a single interaction. Thus, each
entry of our matrix may contain maximally one cost and one benefit term.
Therefore, single altruistic act produces a single pair of cost and benefit.
There is no benefit without a cost. Note that in a similar matrix (\ref
{Donation2}) that ignores role division, we have two costs and two benefits
for the entry describing the interaction between two cooperators. In
addition, in cooperator-noncooperator interactions, the cooperator is always
active and generates the cost and benefit when acting as a row player and a
column player. This is inconsistent with the assumption that the donor pays
a cost, and the receiver consumes the benefit. The proper generalized matrix
of the average payoffs should describe the values averaged over the role
distributions. Let us start from the simplest case of pure frequency
dependence (as in the classical game theory), which implies $
p_{C}^{R}=p_{N}^{R}=g_{C}$. Then the vector of average mortalities $
q^{D}d_{s}^{D}+\left( 1-q^{D}\right) d^{R}(p_{s}^{R})$ resulting from
functions $d_{C}^{f}$ and $d_{N}^{f}$ (\ref{focalC},\ref{focalN}) for both
strategies (which can be decomposed into the background growth rate and the
matrix of differences resulting from the strategies) is

\begin{align*}
& \left[ 
\begin{array}{c}
q^{D}d_{N}^{D}+\left( 1-q^{D}\right) \left[ d^{R}(N)-g_{C}B\right] \\ 
q^{D}\left[ d_{N}^{D}+C\right] +\left( 1-q^{D}\right) \left[ d^{R}(N)-g_{C}B 
\right]
\end{array}
\right] \\
& =\left[ 
\begin{array}{c}
q^{D}d_{N}^{D}+\left( 1-q^{D}\right) d^{R}(N)-\left( 1-q^{D}\right) g_{C}B
\\ 
q^{D}d_{N}^{D}+\left( 1-q^{D}\right) d^{R}(N)-\left( 1-q^{D}\right)
g_{C}B+q^{D}C
\end{array}
\right] \\
& =\left( q^{D}d_{N}^{D}+\left( 1-q^{D}\right) d^{R}(N)\right) \left[ 
\begin{array}{c}
1 \\ 
1
\end{array}
\right] \\
& +\left[ 
\begin{array}{cc}
0 & -\left( 1-q^{D}\right) B \\ 
q^{D}C & q^{D}C-\left( 1-q^{D}\right) B
\end{array}
\right] \left[ 
\begin{array}{c}
1-g_{C} \\ 
g_{C}
\end{array}
\right] .
\end{align*}

Strategy frequencies affect benefits only. Assumption of pairwise
interactions (a single donor helping a single receiver; thus $x=1$) implies $
q^{D}=0.5$, which leads to the matrix equivalent to (\ref{DonationMatrix}),
but with negative benefit and positive cost expressed in terms of
differences in mortality:

\begin{equation}
0.5\left[ 
\begin{array}{cc}
0 & -B \\ 
C & C-B
\end{array}
\right] .
\end{equation}

This is donation game matrix \cite{Panchanathan and Boyd 2003, Marshall 2015}. If we replace pure frequency dependence with assortment mechanism, then
multiplication by the vector $[1-g_{c},g_{c}]^{T}$, should be replaced by
elementwise multiplication by a matrix of assortment probabilities 
\begin{equation*}
\left[ 
\begin{array}{cc}
1-p_{N}^{R} & p_{N}^{R} \\ 
1-p_{C}^{R} & p_{C}^{R}
\end{array}
\right] .
\end{equation*}
Then the average payoff can be obtained by summing the row entries for the
respective strategy. Note that this structure is additive (and has equal
gains from switching property) by definition. Then, for every value of $
p_{C}^{R}$ , the cooperator's payoff has the form $C-p_{C}^{R}B$ and
noncooperators $-p_{N}^{R}B$ (in this case the cost $C$ is paid by a
cooperative donor). Thus, each altruistic act is associated with the same
cost $C$ and benefit $B$, which are equivalent to row effect $x$ and column
effect $y$ in matrix (\ref{matrixEGS}). Nonadditivity (which can be
introduced by adding some value $D$ to some entry of the matrix; Marshall
2015) implies a violation of this property, resulting from the assumption of
passive indistinguishable receivers and the resulting equal cost of helping
for all strategies of receivers. Thus, it is not surprising that Hamilton's
rule is not satisfied for "nonadditive" payoff matrices. When we add a new
parameter $D$\ to the matrix (\ref{DonationMatrix}), then it should also
appear in the resulting condition for cooperation. Then, the standard
Hamilton's rule, which is free from $D$\thinspace\ cannot be valid, and
additional factors such as "synergy coefficients" proposed by Queller \cite{Queller 1985}
should be added. Thus,\ "additivity" is not a limitation of the
counterfactual method but a necessary property resulting from the underlying
biological assumptions and the logic of the modeled class of problems.\ This
is the logical result of the distinction between passive indistinguishable
receivers (whose strategies are latent) and active donors, which bear the
unified cost resulting from the external threat affecting endangered
receivers. This method also works for matrix games with additive payoffs,
but this is a byproduct resulting from the coincidence. Thus, there is no
biological reason that the survival of the cooperative receiver should be
greater than that of a noncooperative receiver, as in the case of
"nonadditive" payoffs \cite{Marshall 2015}. Therefore, while the
regression method can be used in nonadditive models, it seems that for the
counterfactual method, nonadditivity is also not a problem, but for
different reasons. Simply, we don't need "nonadditive" payoffs in the
counterfactual method models.

\section*{Appendix 5: Derivation of the role switching dynamics}

Equations (\ref{malth1}) can be presented in the form 
\begin{eqnarray*}
\dot{n}_{s}^{1} &=&n_{s}^{1}\left[ R_{b}+R_{s}^{1}+\dfrac{n_{s}^{2}}{
n_{s}^{1}}\Lambda ^{2}-\Lambda ^{1}\right] \\
\dot{n}_{s}^{2} &=&n_{s}^{2}\left[ R_{b}+R_{s}^{2}+\dfrac{n_{s}^{1}}{
n_{s}^{2}}\Lambda ^{1}-\Lambda ^{2}\right] .
\end{eqnarray*}
The above system can be rescaled to single replicator equation for two
strategies 
\begin{equation}
\dot{q}_{s}^{1}=q_{s}^{1}(1-q_{s}^{1})\left[ M_{s}^{1}-M_{s}^{2}\right] ,
\end{equation}
where 
\begin{equation*}
M_{s}^{1}=R_{b}+R_{s}^{1}+\dfrac{n_{s}^{2}}{n_{s}^{1}}\Lambda ^{2}-\Lambda
^{1}\text{ \ and \ }M_{s}^{2}=R_{b}+R_{s}^{2}+\dfrac{n_{s}^{1}}{n_{s}^{2}}
\Lambda ^{1}-\Lambda ^{2}.
\end{equation*}
Then, the background growth rate $R_{b}$ cancels out. The terms describing
the switching dynamics in (\ref{malth1}) expressed in terms of frequencies $
q_{s}^{i}=n_{s}^{i}/\left( n_{s}^{1}+n_{s}^{2}\right) $ have the forms 
\begin{eqnarray}
\dfrac{n_{s}^{2}}{n_{s}^{1}}\Lambda ^{2}-\Lambda ^{1} &=&\dfrac{\left(
1-q_{s}^{1}\right) }{q_{s}^{1}}\Lambda ^{2}-\Lambda ^{1}\text{.}
\label{switchingterm} \\
\dfrac{n_{s}^{1}}{n_{s}^{2}}\Lambda ^{1}-\Lambda ^{2} &=&\dfrac{q_{s}^{1}}{
\left( 1-q_{s}^{1}\right) }\Lambda ^{1}-\Lambda ^{2} .
\end{eqnarray}
The separate external bracketed term describing the switching dynamics is:

\begin{align}
q_{s}^{1}&\left( 1-q_{s}^{1}\right) \left( \left[ \dfrac{\left(
1-q_{s}^{1}\right) }{q_{s}^{1}}\Lambda ^{2}-\Lambda ^{1}\right] -\left[ 
\dfrac{q_{s}^{1}}{\left( 1-q_{s}^{1}\right) }\Lambda ^{1}-\Lambda ^{2}\right]
\right) \\
&{}=\left( \left( 1-q_{s}^{1}\right) \left[ \left( 1-q_{s}^{1}\right)
\Lambda ^{2}-q_{s}^{1}\Lambda ^{1}\right] +q_{s}^{1}\left[ \left(
1-q_{s}^{1}\right) \Lambda ^{2}-q_{s}^{1}\Lambda ^{1}\right] \right)  \notag
\\
&{}=\left( 1-q_{s}^{1}\right) \Lambda _{s}^{2}-q_{s}^{1}\Lambda _{s}^{1}.
\end{align}

Therefore, the equation describing the dynamics of the distribution of roles
is: 
\begin{equation}
\dot{q}_{s}^{1}=q_{s}^{1}(1-q_{s}^{1})\left[ R_{s}^{1}-R_{s}^{2}\right] + 
\left[ \left( 1-q_{s}^{1}\right) \Lambda ^{2}-q_{s}^{1}\Lambda ^{1}\right] .
\end{equation}

\section*{Appendix 6: Derivation of the donor/receiver role switching
dynamics}

In effect, we obtain the following system of growth equations: 
\begin{align}
\dot{n}_{C}^{D}& =n_{C}^{D}\left( R_{b}-\left( d_{N}^{D}+C\right) +\left( 
\dfrac{n_{C}^{R}}{n_{C}^{D}}\Lambda ^{R}-\Lambda ^{D}\right) \right) \\
\dot{n}_{N}^{D}& =n_{N}^{D}\left( R_{b}-d_{N}^{D}+\left( \dfrac{n_{N}^{R}}{
n_{N}^{D}}\Lambda ^{R}-\Lambda ^{D}\right) \right) \\
\dot{n}_{C}^{R}& =n_{C}^{R}\left( R_{b}-d_{C}^{R}+\left( \dfrac{n_{C}^{D}}{
n_{C}^{R}}\Lambda ^{D}-\Lambda ^{R}\right) \right)  \label{nrc} \\
\dot{n}_{N}^{R}& =n_{N}^{R}\left( R_{b}-d_{N}^{R}+\left( \dfrac{n_{N}^{D}}{
n_{N}^{R}}\Lambda ^{D}-\Lambda ^{R}\right) \right) ,  \label{nrn}
\end{align}
and after substitution of $d_{C}^{R}$ (\ref{mortC}) and $d_{N}^{R}$ (\ref
{mortN}), equations (\ref{nrc}) and (\ref{nrn}) take the form 
\begin{align}
\dot{n}_{C}^{R}& =n_{C}^{R}\left( R_{b}-d^{R}(N)+p_{C}^{R}B+\left( \dfrac{
n_{C}^{D}}{n_{C}^{R}}\Lambda ^{D}-\Lambda ^{R}\right) \right) \\
\dot{n}_{N}^{R}& =n_{N}^{R}\left( R_{b}-d^{R}(N)+p_{N}^{R}B+\left( \dfrac{
n_{N}^{D}}{n_{N}^{R}}\Lambda ^{D}-\Lambda ^{R}\right) \right) .
\end{align}
We can use (\ref{2strategy}) to describe the switching dynamics (payoff
bracket is negative since it contains only mortalities): 
\begin{equation}
\dot{q}_{s}^{D}=\left( \left( 1-q_{s}^{D}\right) \Lambda
^{R}-q_{s}^{D}\Lambda ^{D}\right) -q_{s}^{D}\left( 1-q_{s}^{D}\right) \left[
d_{s}^{D}(g,q)-d_{s}^{R}(g,q)\right] ,
\end{equation}
leading to 
\begin{align}
\dot{q}_{C}^{D}={}& \left( \left( 1-q_{C}^{D}\right) \Lambda
^{R}-q_{C}^{D}\Lambda ^{D}\right)  \notag \\
& {}-q_{C}^{D}\left( 1-q_{C}^{D}\right) \left[ d_{N}^{D}+C-\left(
d^{R}(N)-p_{C}^{R}B\right) \right] ,
\end{align}
\begin{align}
\dot{q}_{N}^{D}={}& \left( \left( 1-q_{N}^{D}\right) \Lambda
^{R}-q_{N}^{D}\Lambda ^{D}\right)  \notag \\
{}& -q_{N}^{D}\left( 1-q_{N}^{D}\right) \left[ d_{N}^{D}-\left(
d^{R}(N)-p_{N}^{R}B\right) \right] .
\end{align}

\section*{Appendix 7: Derivation of selection dynamics}

Recall the receiver mortalities and more complex functions (\ref{mortC}), ( 
\ref{mortN}), (\ref{focalC}) and (\ref{focalN}). 
\begin{eqnarray}
d_{N}^{D} &=&d_{N}^{D} \\
d_{C}^{D} &=&d_{N}^{D}+C \\
d_{N}^{R} &=&d^{R}(N)-p_{N}^{R}B \\
d_{C}^{R} &=&d^{R}(N)-p_{C}^{R}B
\end{eqnarray}

and the average mortalities: 
\begin{eqnarray}
d_{N}^{f} &=&q_{N}^{D}d_{N}^{D}+\left( 1-q_{N}^{D}\right) d_{N}^{R}  \notag
\\
&=&q_{N}^{D}d_{N}^{D}+\left( 1-q_{N}^{D}\right) \left(
d^{R}(N)-p_{N}^{R}B\right) \\
d_{C}^{f} &=&q_{C}^{D}d_{C}^{D}+\left( 1-q_{C}^{D}\right) d_{C}^{R}  \notag
\\
&=&q_{C}^{D}\left( d_{N}^{D}+C\right) +\left( 1-q_{C}^{D}\right) \left(
d^{R}(N)-p_{C}^{R}B\right) .
\end{eqnarray}
The selection of the strategies will be described by the equation 
\begin{equation}
\dot{g}_{C}=g_{C}\left( 1-g_{C}\right) \left(
d_{N}^{f}(q_{N}^{D})-d_{C}^{f}(q_{N}^{D})\right) .  \label{gene}
\end{equation}
We have bracketed term from (\ref{genedyn}) $(R_{C}-R_{N})=\left(d_{N}^{f}(g,q)-d_{C}^{f}(g,q)\right) $ since mortalities are negative. Let us derive this term, where 
\begin{eqnarray}
d_{C}^{f}(q_{C}^{D}) &=&q_{C}^{D}d_{N}^{D}+\left( 1-q_{C}^{D}\right)
d^{R}(N)-\left( 1-q_{C}^{D}\right) p_{C}^{R}B+q_{C}^{D}C  \label{mortalityC}
\\
d_{N}^{f}(q_{N}^{D}) &=&q_{N}^{D}d_{N}^{D}+\left( 1-q_{N}^{D}\right)
d^{R}(N)-\left( 1-q_{N}^{D}\right) p_{N}^{R}B.  \label{mortalityN}
\end{eqnarray}

The above payoffs can be presented as 
\begin{equation}
d_{C}^{f}(q_{C}^{D})=\tilde{d}_{C}-\left( 1-q_{C}^{D}\right)
p_{C}^{R}B+q_{C}^{D}C  \label{dc}
\end{equation}
and 
\begin{equation}
d_{N}^{f}(q_{N}^{D})=\tilde{d}_{N}-\left( 1-q_{N}^{D}\right) p_{N}^{R}B,
\label{dn}
\end{equation}
where 
\begin{align}
\tilde{d}_{C}& =q_{C}^{D}d_{N}^{D}+\left( 1-q_{C}^{D}\right) d^{R}(N)
\label{basalC} \\
\tilde{d}_{N}& =q_{N}^{D}d_{N}^{D}+\left( 1-q_{N}^{D}\right) d^{R}(N)
\label{basalN}
\end{align}
describe the different basal average mortalities (in addition to the impact
of strategic parameters $C$ and $B$) caused by distributions of states for
both strategies and \ 
\begin{align}
\tilde{d}_{N}-\tilde{d}_{C}={}& \left( q_{N}^{D}-q_{C}^{D}\right) d_{N}^{D}+ 
\left[ \left( 1-q_{N}^{D}\right) -\left( 1-q_{C}^{D}\right) \right] d^{R}(N)
\notag \\
={}& \left( q_{N}^{D}-q_{C}^{D}\right) \left( d_{N}^{D}-d^{R}(N)\right) ,
\label{dn-dc}
\end{align}
and thus, 
\begin{gather}
d_{N}^{f}(q_{N}^{D})-d_{C}^{f}(q_{C}^{D})=  \notag \\
\tilde{d}_{N}-\tilde{d}_{C}-\left( 1-q_{N}^{D}\right) p_{N}^{R}B+\left(
1-q_{C}^{D}\right) p_{C}^{R}B-q_{C}^{D}C=  \notag \\
\left( q_{N}^{D}-q_{C}^{D}\right) \left( d_{N}^{D}-d^{R}(N)\right) +\left[
\left( 1-q_{C}^{D}\right) p_{C}^{R}-\left( 1-q_{N}^{D}\right) p_{N}^{R} 
\right] B-q_{C}^{D}C,  \label{genegrowth}
\end{gather}
leading to the equation on strategy selection (\ref{genedyn}) 
\begin{align}
\dot{g}_{C}=& {}g_{C}\left( 1-g_{C}\right) \left[ \left(
q_{N}^{D}-q_{C}^{D}\right) \left( d_{N}^{D}-d^{R}(N)\right) \right.  \notag
\\
& {}\left. +\left[ \left( 1-q_{C}^{D}\right) p_{C}^{R}-\left(
1-q_{N}^{D}\right) p_{N}^{R}\right] B-q_{C}^{D}C\right] .
\end{align}

\section*{Appendix 8: Derivation of the rule for cooperation\protect\bigskip}

a) Derivation of mortality functions (\ref{mortalityC}) and (\ref{mortalityN}) in terms of parameters $B$, $C$ and $D$

\begin{eqnarray}
d_{C}^{f} &=&q_{C}^{D}d_{N}^{D}+\left( 1-q_{C}^{D}\right) d^{R}(N)-\left(
1-q_{C}^{D}\right) p_{C}^{R}B+q_{C}^{D}C \\
&=&q_{C}^{D}d_{N}^{D}+\left( 1-q_{C}^{D}\right) \left(
d_{N}^{D}+B+C+D\right) -\left( 1-q_{C}^{D}\right) p_{C}^{R}B+q_{C}^{D}C \\
&=&d_{N}^{D}+\left( 1-q_{C}^{D}\right) (\left( 1-p_{C}^{R}\right) B+D)+C \\
&& \\
d_{N}^{f} &=&q_{N}^{D}d_{N}^{D}+\left( 1-q_{N}^{D}\right) d^{R}(N)-\left(
1-q_{N}^{D}\right) p_{N}^{R}B \\
&=&q_{N}^{D}d_{N}^{D}+\left( 1-q_{N}^{D}\right) \left(
d_{N}^{D}+B+C+D\right) -\left( 1-q_{N}^{D}\right) p_{N}^{R}B \\
&=&d_{N}^{D}+\left( 1-q_{N}^{D}\right) \left( C+D\right) +\left(
1-q_{N}^{D}\right) \left( 1-p_{N}^{R}\right) B \\
&=&d_{N}^{D}+\left( 1-q_{N}^{D}\right) \left( \left( 1-p_{N}^{R}\right)
B+D+C\right) .
\end{eqnarray}
b) Derivation of the rule for cooperation

Condition $d_{N}^{f}>d_{C}^{f}$ is

\begin{equation}
\left( 1-q_{N}^{D}\right) \left( \left( 1-p_{N}^{R}\right) B+D+C\right)
>\left( 1-q_{C}^{D}\right) \left( \left( 1-p_{C}^{R}\right) B+D\right) +C
\end{equation}
\begin{align}
\left( 1-q_{N}^{D}\right) \left( \left( 1-p_{N}^{R}\right) B+D\right)
-\left( 1-q_{C}^{D}\right) (\left( 1-p_{C}^{R}\right) B+D)& >C-\left(
1-q_{N}^{D}\right) C \\
\left[ \left( 1-q_{N}^{D}\right) \left( 1-p_{N}^{R}\right) -\left(
1-q_{C}^{D}\right) \left( 1-p_{C}^{R}\right) \right] B+\left[ \left(
1-q_{N}^{D}\right) -\left( 1-q_{C}^{D}\right) \right] D& >q_{N}^{D}C \\
\left[ \left( 1-q_{N}^{D}\right) \left( 1-p_{N}^{R}\right) -\left(
1-q_{C}^{D}\right) \left( 1-p_{C}^{R}\right) \right] B+\left[
q_{C}^{D}-q_{N}^{D}\right] D& >q_{N}^{D}C .
\end{align}
c) Derivation of the relative fitness effect surfaces by substitution $
D=1-B-C$ 
\begin{eqnarray*}
d_{C}^{f} &=&\left( 1-q_{C}^{D}\right) \left( \left( 1-p_{C}^{R}\right) B+ 
\left[ 1-B-C\right] \right) +C \\
&=&\left( 1-q_{C}^{D}\right) \left( \left( 1-p_{C}^{R}\right) B+1-B\right)
+C-\left( 1-q_{C}^{D}\right) C \\
&=&\left( 1-q_{C}^{D}\right) \left( 1-p_{C}^{R}B\right) +q_{C}^{D}C
\end{eqnarray*}
\begin{eqnarray*}
d_{N}^{f} &=&\left( 1-q_{N}^{D}\right) \left( \left( 1-p_{N}^{R}\right) B+ 
\left[ 1-B-C\right] +C\right) \\
&=&\left( 1-q_{N}^{D}\right) \left( 1-p_{N}^{R}B\right) .
\end{eqnarray*}
\bigskip

\section*{Appendix 9: Substitution of equilibria of the switching dynamics
	to the cooperation rule}

a) Calculation of the stable role distributions for constant switching rates.

Recall the switching dynamics (\ref{q-C}) and (\ref{q-N})\bigskip

\begin{align}
	\dot{q}_{C}^{D}={}& \left( \left( 1-q_{C}^{D}\right) \Lambda
	^{R}-q_{C}^{D}\Lambda ^{D}\right)   \notag \\
	& {}-q_{C}^{D}\left( 1-q_{C}^{D}\right) \left[ d_{N}^{D}+C-\left(
	d^{R}(N)-p_{C}^{R}B\right) \right] 
\end{align}
\begin{align}
	\dot{q}_{N}^{D}={}& \left( \left( 1-q_{N}^{D}\right) \Lambda
	^{R}-q_{N}^{D}\Lambda ^{D}\right)   \notag \\
	{}& -q_{N}^{D}\left( 1-q_{N}^{D}\right) \left[ d_{N}^{D}-\left(
	d^{R}(N)-p_{N}^{R}B\right) \right] .
\end{align}
Recall that $d^{R}(N)=d_{N}^{D}+C+B+D$ and assume for simplicity

\begin{eqnarray*}
	-A_{C} &=&d_{N}^{D}+C-\left( d^{R}(N)-p_{C}^{R}B\right)  \\
	&=&d_{N}^{D}-d^{R}(N)+C+p_{C}^{R}B \\
	&=&-B-D+p_{C}^{R}B \\
	&=&-\left( 1-p_{C}^{R}\right) B-D \\
	&=&-\left[ \left( 1-p_{C}^{R}\right) B+D\right] 
\end{eqnarray*}

\begin{eqnarray*}
	-A_{N} &=&d_{N}^{D}-\left( d^{R}(N)-p_{N}^{R}B\right)  \\
	&=&d_{N}^{D}-d^{R}(N)+p_{N}^{R}B \\
	&=&-C-B-D+p_{N}^{R}B \\
	&=&-\left( 1-p_{N}^{R}\right) B-C-D \\
	&=&-\left[ \left( 1-p_{N}^{R}\right) B+C+D\right] .
\end{eqnarray*}
Then the switching dynamics can be presented in the form

\begin{eqnarray}
	\dot{q}_{C}^{D} &=&\left( \left( 1-q_{C}^{D}\right) \Lambda
	^{R}-q_{C}^{D}\Lambda ^{D}\right) +q_{C}^{D}\left( 1-q_{C}^{D}\right) \left[
	\left( 1-p_{C}^{R}\right) B+D\right] , \\
	\dot{q}_{N}^{D} &=&\left( \left( 1-q_{N}^{D}\right) \Lambda
	^{R}-q_{N}^{D}\Lambda ^{D}\right) +q_{N}^{D}\left( 1-q_{N}^{D}\right) \left[
	\left( 1-p_{N}^{R}\right) B+C+D\right] ,
\end{eqnarray}

Then both equations satisfy the general form where $A_{s}>0$ 
\begin{eqnarray*}
	\dot{q}_{s}^{D} &=&\left( \left( 1-q_{s}^{D}\right) \Lambda
	^{R}-q_{s}^{D}\Lambda ^{D}\right) +q_{s}^{D}\left( 1-q_{s}^{D}\right) A_{s}
	\\
	&=&\Lambda ^{R}-q_{s}^{D}\left( \Lambda ^{R}+\Lambda ^{D}\right) +\left(
	q_{s}^{D}-\left( q_{s}^{D}\right) ^{2}\right) A_{s} \\
	&=&-A_{s}\left( q_{s}^{D}\right) ^{2}-\left[ \Lambda ^{R}+\Lambda ^{D}-A_{s}
	\right] q_{s}^{D}+\Lambda ^{R} .
\end{eqnarray*}

This is the quadratic equation. Note that for $q_{s}^{D}=0$ we have $\dot{q}
_{s}^{D}=\Lambda ^{R}$, and for $q_{s}^{D}=1$ we have $\dot{q}
_{s}^{D}=-\Lambda ^{D}$ This implies that one stable root should exist in
the interior of the unit interval. We have  $\Delta =\left[ \Lambda
^{R}+\Lambda ^{D}-A_{s}\right] ^{2}+4A_{s}\Lambda ^{R}>4A_{s}\Lambda ^{R}>0$
. since all coefficients are positive. Thus for nonzero parameters we have
always two\ roots. From Viete'a formula we have that the product of roots
equals $\Lambda ^{R}/(-A_{s})<0$, thus one root is negative (unstable) and
one is positive.(stable) since r.h.s. of the switching dynamics is positive
between them.The stable point will be

\begin{equation*}
	\tilde{q}_{s}^{D}=\frac{\left[ \Lambda ^{R}+\Lambda ^{D}-A_{s}\right] -\sqrt{
			\left[ \Lambda ^{R}+\Lambda ^{D}-A_{s}\right] ^{2}+4A_{s}\Lambda ^{R}}}{
		-2A_{s}},
\end{equation*}

Let us check the $\tilde{q}_{s}^{D}<1$ condition, which implies:

\begin{gather*}
	\left[ \Lambda ^{R}+\Lambda ^{D}-A_{s}\right] -\sqrt{\left[ \Lambda
		^{R}+\Lambda ^{D}-A_{s}\right] ^{2}+4A_{s}\Lambda ^{R}}>-2A_{s} \\
	\Lambda ^{R}+\Lambda ^{D}+A_{s}>\sqrt{\left[ \Lambda ^{R}+\Lambda ^{D}-A_{s}
		\right] ^{2}+4A_{s}\Lambda ^{R}} \\
	\left[ \Lambda ^{R}+\Lambda ^{D}+A_{s}\right] ^{2}>\left[ \Lambda
	^{R}+\Lambda ^{D}-A_{s}\right] ^{2}+4A_{s}\Lambda ^{R} \\
	2\left[ \Lambda ^{R}+\Lambda ^{D}\right] A_{s}>-2\left[ \Lambda ^{R}+\Lambda
	^{D}\right] A_{s}+4A_{s}\Lambda ^{R} \\
	4\Lambda ^{D}A_{s}>0
\end{gather*}

which is always true since all parameters are positive. Then, the unique
attractors of the switching dynamics for both strategies are

\begin{gather}
	\tilde{q}_{C}^{D}=  \notag \\
	\frac{-\left[ \Lambda ^{R}+\Lambda ^{D}-\left( 1-p_{C}^{R}\right) B-D\right]
		+\sqrt{\left[ \Lambda ^{R}+\Lambda ^{D}-\left( 1-p_{C}^{R}\right) B-D\right]
			^{2}+4\Lambda ^{R}\left[ \left( 1-p_{C}^{R}\right) B+D\right] }}{2\left[
		\left( 1-p_{C}^{R}\right) B+D\right] } \\
	\tilde{q}_{N}^{D}=  \notag \\
	\frac{-\left[ \Lambda ^{R}+\Lambda ^{D}-\left( 1-p_{N}^{R}\right) B-C-D
		\right] +\sqrt{\left[ \Lambda ^{R}+\Lambda ^{D}-\left( 1-p_{N}^{R}\right)
			B-C-D\right] ^{2}+4\Lambda ^{R}\left[ \left( 1-p_{N}^{R}\right) B+C+D\right] 
	}}{2\left[ \left( 1-p_{N}^{R}\right) B+C+D\right] }
\end{gather}
\bigskip $\bigskip $

b) Derivation of the rule for cooperation for constant switching rates.

Recall the rule (\ref{coop2}) 
\begin{equation}
	\left[ \left( 1-q_{N}^{D}\right) \left( 1-p_{N}^{R}\right) -\left(
	1-q_{C}^{D}\right) \left( 1-p_{C}^{R}\right) \right] B+\left[
	q_{C}^{D}-q_{N}^{D}\right] D)>q_{N}^{D}C
\end{equation}
Now we can substitute the roots to the general rule for cooperation (\ref
{coop2}). For simplicity we limit ourselves to the case when\emph{\ }$D=0$ 
\emph{.} To simplify this task we can rearrange the rule (\ref{coop2}):
\bigskip 
\begin{gather*}
	\left[ \left( 1-q_{N}^{D}\right) \left( 1-p_{N}^{R}\right) -\left(
	1-q_{C}^{D}\right) \left( 1-p_{C}^{R}\right) \right] B>q_{N}^{D}C \\
	\left( 1-p_{N}^{R}\right) B-q_{N}^{D}\left( 1-p_{N}^{R}\right)
	B-q_{N}^{D}C-\left( 1-p_{C}^{R}\right) B+q_{C}^{D}\left( 1-p_{C}^{R}\right)
	B>0 \\
	\left( 1-p_{N}^{R}\right) B-\left( 1-p_{C}^{R}\right) B>q_{N}^{D}\left(
	\left( 1-p_{N}^{R}\right) B+C\right) -q_{C}^{D}\left( 1-p_{C}^{R}\right) B \\
	\left[ p_{C}^{R}-p_{N}^{R}\right] B>q_{N}^{D}\left( \left(
	1-p_{N}^{R}\right) B+C\right) -q_{C}^{D}\left( 1-p_{C}^{R}\right) B .
\end{gather*}

After substitution of the $\tilde{q}_{C}^{D}$ and $\tilde{q}_{N}^{D}$ the
rule have form\bigskip

\begin{gather}
	\left[ p_{C}^{R}-p_{N}^{R}\right] B>  \notag \\
	\frac{\sqrt{\left[ \Lambda ^{R}+\Lambda ^{D}-\left( 1-p_{N}^{R}\right) B-C
			\right] ^{2}+4\Lambda ^{R}\left( \left( 1-p_{N}^{R}\right) B+C\right) }-
		\left[ \Lambda ^{R}+\Lambda ^{D}-\left( 1-p_{N}^{R}\right) B-C\right] }{
		2\left( \left( 1-p_{N}^{R}\right) B+C\right) }\left[ \left(
	1-p_{N}^{R}\right) B+C\right]   \notag \\
	-\frac{\sqrt{\left[ \Lambda ^{R}+\Lambda ^{D}-\left( 1-p_{C}^{R}\right) B
			\right] ^{2}+4\Lambda ^{R}\left( \left( 1-p_{C}^{R}\right) B\right) }-\left[
		\Lambda ^{R}+\Lambda ^{D}-\left( 1-p_{C}^{R}\right) B\right] }{2\left(
		\left( 1-p_{C}^{R}\right) B\right) }\left( 1-p_{C}^{R}\right) B .
\end{gather}

\section*{Appendix 10 Derivation of the kin selection case}

Derivation of the fractions of the unhelped individuals 
\begin{eqnarray*}
&&\left( 1-q_{N}^{D}\right) \left( 1-p_{N}^{kin}\right) \\
&=&\left( 1-q_{N}^{D}\right) \left( 1-(1-r)g_{C}\right) \\
&&\text{and} \\
&&\left( 1-q_{C}^{D}\right) \left( 1-p_{C}^{kin}\right) \\
&=&\left( 1-q_{C}^{D}\right) \left( 1-r-(1-r)g_{C}\right) \\
&=&\left( 1-q_{C}^{D}\right) \left( 1-(1-r)g_{C}\right) -\left(
1-q_{C}^{D}\right) r \\
&=&\left( 1-q_{C}^{D}\right) (1-r)\left( 1-g_{C}\right) .
\end{eqnarray*}

Then the bracketed term $\left[ \left( 1-q_{N}^{D}\right) \left(
1-p_{N}^{kin}\right) -\left( 1-q_{C}^{D}\right) \left( 1-p_{C}^{kin}\right) 
\right] $ will be:\bigskip 
\begin{eqnarray*}
&&\left( 1-q_{N}^{D}\right) \left( 1-(1-r)g_{C}\right) -\left(
1-q_{C}^{D}\right) \left( 1-(1-r)g_{C}\right) +\left( 1-q_{C}^{D}\right) r \\
&=&\left[ q_{C}^{D}-q_{N}^{D}\right] \left( 1-(1-r)g_{C}\right) +\left(
1-q_{C}^{D}\right) r \\
&&\text{Bracket revealing the impact of }r\text{ will be} \\
&=&\left[ q_{C}^{D}-q_{N}^{D}\right] \left( 1-g_{C}+rg_{C}\right) +\left(
1-q_{C}^{D}\right) r \\
&=&\left[ q_{C}^{D}-q_{N}^{D}\right] \left( 1-g_{C}\right) +\left[
q_{C}^{D}-q_{N}^{D}\right] rg_{C}+\left( 1-q_{C}^{D}\right) r \\
&=&\left[ q_{C}^{D}-q_{N}^{D}\right] \left( 1-g_{C}\right) +\left( \left[
q_{C}^{D}-q_{N}^{D}\right] g_{C}+\left( 1-q_{C}^{D}\right) \right) r \\
&=&\left[ q_{C}^{D}-q_{N}^{D}\right] \left( 1-g_{C}\right) +\left( \left(
g_{C}-1\right) q_{C}^{D}+1-q_{N}^{D}g_{C}\right) r \\
&=&\left[ q_{C}^{D}-q_{N}^{D}\right] \left( 1-g_{C}\right) +\left( 1-\left[
\left( 1-g_{C}\right) q_{C}^{D}+g_{C}q_{N}^{D}\right] \right) r .
\end{eqnarray*}

\section*{Appendix 11: Proof of Theorem 1}

For $\left[ q_{C}^{D}-q_{N}^{D}\right] >0$ formula (\ref{kinhamD0}) is
satisfied when\bigskip

\begin{eqnarray*}
\left[ q_{C}^{D}-q_{N}^{D}\right] \left( 1-(1-r)g_{C}\right) +\left(
1-q_{C}^{D}\right) r &>&\frac{q_{N}^{D}C}{B} ,\\
1-(1-r)g_{C} &>&\frac{q_{N}^{D}\frac{C}{B}-\left( 1-q_{C}^{D}\right) r}{
q_{C}^{D}-q_{N}^{D}} ,\\
1-\frac{q_{N}^{D}\frac{C}{B}-\left( 1-q_{C}^{D}\right) r}{
q_{C}^{D}-q_{N}^{D} } &>&(1-r)g_{C} ,\\
\frac{1-\frac{q_{N}^{D}\frac{C}{B}-\left( 1-q_{C}^{D}\right) r}{
q_{C}^{D}-q_{N}^{D}}}{(1-r)} &>&g_{C} ,\\
g_{C} &<&\frac{q_{C}^{D}-q_{N}^{D}-q_{N}^{D}\frac{C}{B}+\left(
1-q_{C}^{D}\right) r}{\left( q_{C}^{D}-q_{N}^{D}\right) (1-r)} ,\\
g_{C} &<&\tilde{g}_{C}=\frac{q_{C}^{D}+\left( 1-q_{C}^{D}\right)
r-q_{N}^{D}\left( 1+\frac{C}{B}\right) }{\left( q_{C}^{D}-q_{N}^{D}\right)
(1-r)} .
\end{eqnarray*}

For $\left( q_{C}^{D}-q_{N}^{D}\right) >0$ we need positive numerator for
positive values of $\tilde{g}_{C}$ for which the above inequality will be
satified: 
\begin{eqnarray*}
q_{C}^{D}+\left( 1-q_{C}^{D}\right) r-q_{N}^{D}\left( 1+\frac{C}{B}\right)
&>&0 ,\\
q_{N}^{D} &<&\frac{q_{C}^{D}+\left( 1-q_{C}^{D}\right) r}{\left( 1+\frac{C}{
B }\right) }
\end{eqnarray*}

and for $\tilde{g}_{C}<1$ 
\begin{eqnarray*}
q_{C}^{D}+\left( 1-q_{C}^{D}\right) r-q_{N}^{D}\left( 1+\frac{C}{B}\right)
&<&\left( q_{C}^{D}-q_{N}^{D}\right) (1-r) ,\\
q_{C}^{D}-q_{C}^{D}(1-r)+\left( 1-q_{C}^{D}\right) r-q_{N}^{D}\left( 1+\frac{
C}{B}\right) +q_{N}^{D}(1-r) &<&0 ,\\
q_{C}^{D}r+\left( 1-q_{C}^{D}\right) r-q_{N}^{D}\left( r+\frac{C}{B}\right)
&<&0 ,\\
r-q_{N}^{D}\left( r+\frac{C}{B}\right) &<&0 ,\\
r &<&q_{N}^{D}\left( r+\frac{C}{B}\right) ,\\
q_{N}^{D} &>&\frac{r}{r+\frac{C}{B}} .
\end{eqnarray*}

Then for $0<\tilde{g}_{C}<1$ we need 
\begin{equation*}
\frac{q_{C}^{D}+\left( 1-q_{C}^{D}\right) r}{\left( 1+\frac{C}{B}\right) }
>q_{N}^{D}>\frac{r}{r+\frac{C}{B}} .
\end{equation*}

Let us check the inequality 
\begin{eqnarray*}
\frac{q_{C}^{D}+\left( 1-q_{C}^{D}\right) r}{\left( 1+\frac{C}{B}\right) }
&>&\frac{r}{r+\frac{C}{B}} ,\\
\frac{q_{C}^{D}(1-r)+r}{\left( 1+\frac{C}{B}\right) } &>&\frac{r}{r+\frac{C}{
B}} ,\\
q_{C}^{D}(1-r)+r &>&\frac{r}{r+\frac{C}{B}}\left( 1+\frac{C}{B}\right) ,\\
q_{C}^{D}(1-r) &>&\frac{r}{r+\frac{C}{B}}\left( 1+\frac{C}{B}\right) -r  ,\\
q_{C}^{D} &>&\frac{r}{(1-r)}\left( \frac{1+\frac{C}{B}}{r+\frac{C}{B}}
-1\right)  ,\\
q_{C}^{D} &>&\frac{r}{(1-r)}\left( \frac{1-r}{r+\frac{C}{B}}\right) ,\\
q_{C}^{D} &>&\frac{r}{r+\frac{C}{B}} .
\end{eqnarray*}

For $\left[ q_{C}^{D}-q_{N}^{D}\right] <0$ we have exactly the same
derivation but with the opposite sign of inequality. End of the proof.

\end{document}